\documentclass[journal]{IEEEtran}
\usepackage{epsfig,amsmath,amssymb,epsf,cite,scalefnt,graphicx,subfig,array} 
\usepackage{cases,epstopdf,grffile}

\def\bU{{\mathbf{U}}}    
\def\bZ{{\mathbf{Z}}}

 \def\ibb{{\pmb{b}}}   
  \def\ibh{{\pmb{h}}}

\def\ibu{{\pmb{u}}} \def\ibv{{\pmb{v}}} \def\ibw{{\pmb{w}}} \def\ibx{{\pmb{x}}} 
\def\ibz{{\pmb{z}}}

   \def\ibI{{\pmb{I}}}

\def\span{\mathop{\mathrm{span}}}

\renewcommand{\baselinestretch}{1.0}

     \def\d4{\!\!\!\!}              \def\eps{\epsilon}

 \def\bPhi{\mathbf{\Phi}}   \def\lam{\lambda} 
 \def\blam{\boldsymbol{\lambda}}
\def\beps{\boldsymbol{\epsilon}}

  \def\-{\! - \!}  \def\+{\! + \!}  \def\={\! = \!}  \def\>{\! > \!}

\newcommand{\bef}{\begin{figure}}
\newcommand{\eef}{\end{figure}}
\newcommand{\beq}{\begin{eqnarray}}
\newcommand{\eeq}{\end{eqnarray}}

\newcommand{\qed}{\nobreak \ifvmode \relax \else
\ifdim\lastskip<1.5em \hskip-\lastskip \hskip1.5em plus0em
minus0.5em \fi \nobreak \vrule height0.5em width0.5em
depth0.25em\fi}

\ifCLASSINFOpdf
\else
\fi

\begin{document}

\title{Fast $L_1$-Minimization Algorithm for Sparse Approximation Based on an Improved LPNN-LCA framework}

\author{Hao~Wang,
        Ruibin~Feng,
        and~Chi-Sing~Leung,~\IEEEmembership{Member,~IEEE,}
\thanks{Hao~Wang, Ruibin~Feng, and Chi-Sing~Leung are with the Department of Electronic Engineering, City University of Hong Kong, Kowloon Tong, Hong Kong, China.}}

\markboth{IEEE TRANSACTIONS ON SIGNAL PROCESSING,~Vol.~1, No.~1, September~2018}%
{Shell \MakeLowercase{\textit{et al.}}: Fast $L_1$-Minimization Algorithm for Sparse Approximation Based on an Improved LPNN-LCA framework}

\maketitle

\begin{abstract}
The aim of sparse approximation is to estimate a sparse signal according to the measurement matrix and an observation vector. It is widely used in data analytics, image processing, and communication, etc.
Up to now, a lot of research has been done in this area, and many off-the-shelf algorithms have been proposed.
However, most of them cannot offer a real-time solution.
To some extent, this shortcoming limits its application prospects. To address this issue, we devise a novel sparse approximation algorithm based on Lagrange programming neural network (LPNN), locally competitive algorithm (LCA), and projection theorem.
LPNN and LCA are both analog neural network which can help us get a real-time solution. The non-differentiable objective function can be solved by the concept of LCA. Utilizing the projection theorem, we further modify the dynamics and proposed a new system with global asymptotic stability. Simulation results show that the proposed sparse approximation method has the real-time solutions with satisfactory MSEs.

\end{abstract}


%
\IEEEpeerreviewmaketitle

\section{Introduction}\label{section1}

%
%
%
Sparse approximation is frequently used in many different application areas including data analytic, image processing, communication, feature extraction, audio processing and so on \cite{wright2009robust,bach2010sparse,huang2006sparse,li2004analysis,wright2010sparse}. The aim of sparse approximation algorithms is to represent an observation using a sparse signal selected from a specified measurement matrix. The relationship between the unknown sparse signal and the observation can be described as
\beq
\bPhi\ibx=\ibb,
\label{eq-0.1}
\eeq
where $\ibx\in\mathbb{R}^n$ is an unknown sparse vector, $\ibb\in\mathbb{R}^m$ is an observation vector, and $\bPhi\in\mathbb{R}^{m\times n}$ is the measurement matrix (also known as dictionary) with rank $m$ ($m<n$). The matrix $\bPhi$ is fat with full row rank. Apparently, an infinite number of solutions are available for this equality. Therefore, constraints must be introduced. In sparse approximation, the number of nonzero elements in $\ibx$ is denoted by $t$, the solution with the smallest $t$ is the best representation. Because for real data, like communication signals and images, even though their observations are in high-dimensional spaces, the actual signals are organized in some lower-dimensional subspaces, i.e., $t<m$.  Hence the spare representation problem is to solve:
\begin{subequations}\label{eq-0.2}
\beq
\min \,\,&||\ibx||_0 \\
\textbf{s.t.}\,\, &\bPhi\ibx=\ibb,
\eeq
\end{subequations}
where $||\cdot||_0$ is the so-called $l_0$-norm. $||\ibx||_0$ implies the number of nonzero elements in $\ibx$. The problem in \eqref{eq-0.2} is a non-convex.
And, due to $l_0$-norm, it is NP-hard~\cite{donoho2006most}. Hence, in the first kind of methods, approximate functions are used to replace the $||\ibx||_0$ term.
It is well known that $l_1$-norm is the best convex approximate function of $l_0$-norm. Thus the original problem in \eqref{eq-0.2} can be rewritten as
\begin{subequations}\label{eq-bp}
\beq
\min \,\,&||\ibx||_1 \\
\textbf{s.t.}\,\, &\bPhi\ibx=\ibb,
\eeq
\end{subequations}
which is known as basis pursuit (BP).
It can be also transformed into the unconstrained form given by
\beq\label{eq-lasso}
\min \,\, \frac{1}{2} \| \ibb -\bPhi \ibx \|_2^2  + \kappa \| \ibx \|_1,
\eeq
where $\kappa\in\mathbb{R}$ is a trade-off parameter.  \eqref{eq-lasso} is known as least absolute shrinkage and selection operator (LASSO) problem.
Both \eqref{eq-bp} and \eqref{eq-lasso} can be used to calculate the original solution of problem~\eqref{eq-0.2} under certain hypotheses \cite{donoho2006most}. And, for solving the problem in \eqref{eq-bp} and \eqref{eq-lasso}, some elegant implementation packages, for example L1Magic \cite{candes2005l1} and SPGL1 \cite{BergFriedlander2008,spgl12007}, are available. Comparing \eqref{eq-bp} and \eqref{eq-lasso}, \eqref{eq-bp} is more attractive. Because it does not need any trade-off parameter and the solution of this model normally has better performance. Therefore, the proposed algorithm also uses the BP model in \eqref{eq-bp}.

Another common type of methods are greedy iterative algorithms which include the matching pursuit (MP) algorithm \cite{mallat1993matching} and its variants. The basic procedures of this kind of algorithms are shown as follows. First, the column in $\bPhi$ which best matches with $\ibb$ is found. Second, the residual is computed. Third, the column in $\bPhi$ which best matches with the residual is calculated; Finally, we repeat the second and third step several times and then calculate a linear summation of all selected columns. Theoretically, the linear summation of these columns is the most optimal sparse representation of the observation.

In addition to these two kinds of algorithms, there are several other methods including gradient descent based methods \cite{nowak2007gradient}, Dantzig selector method\cite{candes2007dantzig}, etc.


In practice, the computational speed is a major barrier to the real-time signal processing with high-dimensional signals \cite{balavoine2012convergence}. However, existing sparse approximation algorithms normally suffer from one or more the following drawbacks:
(i) Computational speed for solving the sparse approximation problem largely depends on the dimension and denseness of the problem. With the increasing of dimensions and denseness, the performance of these conventional methods is worse and worse.
(ii) Most of them cannot generate exactly zero-valued coefficients in finite time \cite{rozell2008sparse}.
(iii) Generally speaking, they can only calculate a suboptimal solution due to the relaxation.

The analog neural network is very attractive when real-time solution are needed \cite{cochocki1993neural,hopfield1982neural,chua1984nonlinear}. Because the analog neural network can be implemented by hardware circuit and its parallel architecture can eliminate the influence of high dimensions and great denseness.
Hence, in this paper, we use the analog neural network to solve the BP problem.
LPNN is an analog neural network which is generally used to solve the nonlinear constrained optimization problems \cite{zhang1992lagrange}. However, LPNN requires that all its objective function and constraints should be twice differentiable. Obviously, BP problem does not satisfy this requirement. Hence it cannot be directly used to solve the BP problem.
LCA is another analog neural network, which can be used for solving the sparse approximation problem in \eqref{eq-lasso}. Even though it is effective to handle the non-differentiable term, this method can only solve the unconstrained optimization problem. For solving the BP problem with LCA, we have to introduce an appropriate trade-off parameter and transform the original problem into the unconstrained form. Besides, LCA cannot ensure that the original constraints in BP are satisfied.
Hence, in our previous work, LPNN and LCA are combined to solve the BP problem. Even though the performance of this method is superior, the global stability of this system is hard to prove.
Inspired by the projection neural network \cite{xia2002projection,xia2004general},
we note that the dynamics of LPNN-LCA method also contain a projection operation to a convex set, and their structure can be further modified. After the modification, its equilibrium point is still equivalent to the optimal solution of the original BP problem, and the whole system has global stability in the sense of Lyapunov.

%

The rest of paper is organized as follows.  In Section~\ref{section2}, the background of LPNN, LCA and projection neural network are described. In Section~\ref{section3}, the proposed algorithm are developed. The global convergence of our algorithm is proved in Section~\ref{section4}. Experimental results for algorithm evaluation and comparison are provided in Section~\ref{section5}. Finally, conclusions are drawn in Section~\ref{section6}.

\section{Background}\label{section2}
\subsection{Notation}
We use a lower-case or upper-case letter to represent a scalar while vectors and matrices are denoted by bold lower-case and upper-case letters, respectively. The transpose operator is denoted as $(\centerdot)^ \mathrm{T}$, and $\ibI$ and $\mathbf{0}$ respectively represent the identity matrix and zero matrix with appropriate dimensions. Other mathematical symbols are defined in their first appearance.

\subsection{Lagrange Programming Neural Network}
LPNN is an analog neural network, which can be used to solve a general nonlinear constrained optimization problem given by
\begin{subequations}\label{eq-1.6}
\beq
\min\limits_{\ibx} &\,\,f(\ibx) \\
\mbox{s.t.}&\,\, \ibh(\ibx)=0.
\eeq
\end{subequations}
where $\ibx=[x_1,\cdots,x_n]^\mathrm{T}$ is the variable vector, $f:\mathbb{R}^n \to \mathbb{R}$ is the objective function, $\ibh:\mathbb{R}^n \to \mathbb{R}^m$ ($m<n$) represents $m$ equality constraints. In LPNN, we first set up its Lagrangian:
\beq
\label{eq-1.7}
L(\ibx,\blam)=f(\ibx)+\blam^\mathrm{T}\ibh(\ibx)
\eeq
where $\blam=[\lam_1,\cdots,\lam_m]^\mathrm{T}$ is the Lagrange multiplier.
There are two kinds of neurons in LPNN, namely, variable neurons and Lagrangian neurons.
The $n$ variable neurons are used to hold the decision variable vector $\ibx$ while
the $m$ Lagrangian neurons hold the Lagrange multiplier vector $\blam$.
In the LPNN framework, the dynamics of the neurons are
\begin{subequations}\label{eq-1.8}
\beq
\epsilon\frac{d\ibx}{dt}=& \displaystyle -\frac{\partial L(\ibx,\blam)}{\partial \ibx}, \\
\epsilon\frac{d\blam}{dt}=& \displaystyle \frac{\partial L(\ibx,\blam)}{\partial \blam},
\eeq
\end{subequations}
where $\epsilon$ is the time constant of the circuit. Without loss of generality, we let $\epsilon=1$ in this paper.
The differential equations in (\ref{eq-1.8}) govern the state transition of the neurons.
After the neurons settle down at an equilibria, the solution is obtained by measuring the neuron outputs at the equilibrium point. The purpose of (\ref{eq-1.8}a) is to seek for a state with the minimum objective value, while (\ref{eq-1.8}b) aims to constrain its outputs into the feasible region.
With (\ref{eq-1.8}), the network will settle down at a stable state if several mild conditions are satisfied~\cite{zhang1992lagrange,lpnn3,lpnn4}.
It is worth noting that both $f$ and $\ibh$ should be differentiable. Otherwise, the dynamics cannot be defined. Thus, due to the $l_1$-norm term in \eqref{eq-bp}, it cannot be directly solved by LPNN.

\subsection{Locally Competitive Algorithm}
The LCA~\cite{rozell2008sparse} is also an analog neural network which is devised for solving the unconstrained sparse approximation problem in~\eqref{eq-lasso}.

\begin{figure}[htb]
\centering
\centerline{\includegraphics[width=2.5in]{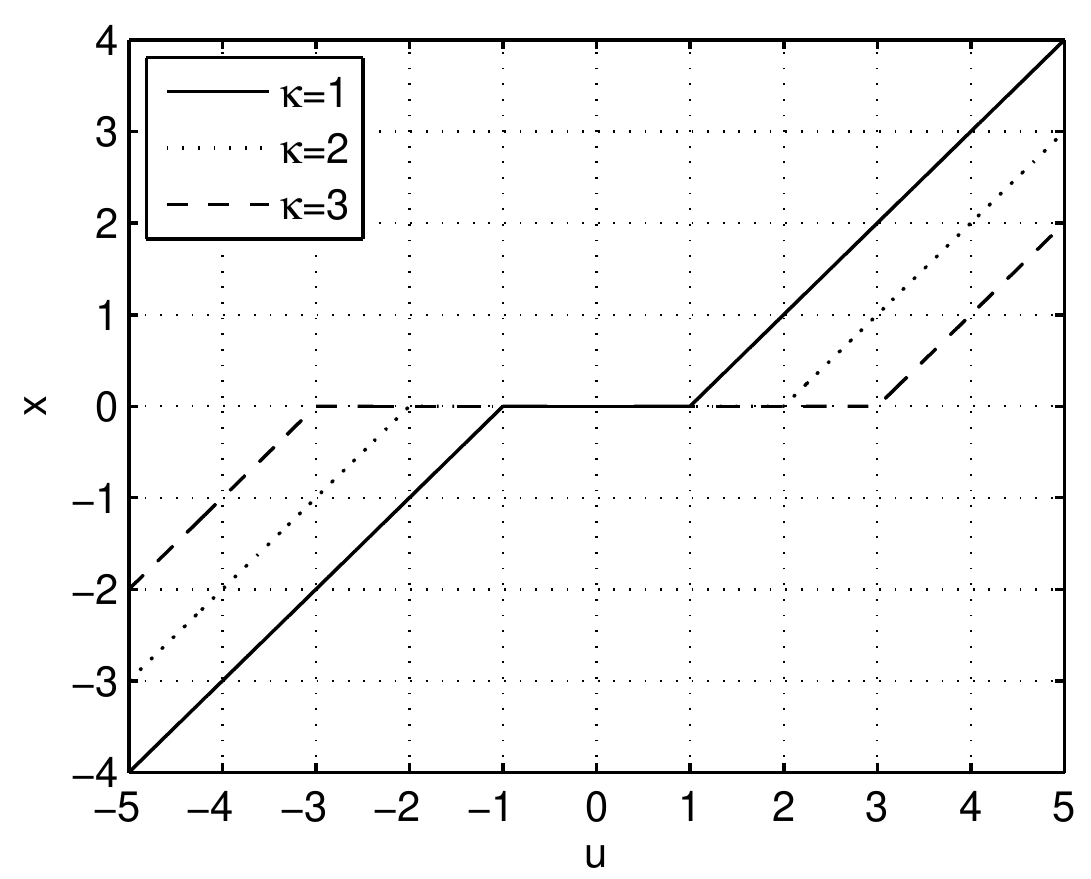}}
\caption{General threshold function.}
\label{threshold}
\end{figure}

Due to the non-differentiable term $\kappa \| \ibx \|_1$, LCA introduces an internal state variable $\ibu=[u_1,\cdots,u_n]^\mathrm{T} $ and defines an element-wise relationship between $\ibx$ and $\ibu$ which is given by
\begin{equation}
x_i = T_{\kappa}(u_i) = \left\{ \begin{array}{lcl}
0, &  |u_i| \leq \kappa, \\
u_i - \kappa \mbox{sign} (u_i),  & |u_i| > \kappa,
\end{array}\right.
\label{internal2}
\end{equation}
Where $x_i$ and $u_i$ denote the i-th element of $\ibx$ and $\ibu$ respectively.
The shapes of function \eqref{internal2} with $\kappa=1$, $\kappa=2$ and $\kappa=3$ are shown in Fig.~\ref{threshold}. It is observed that the $\kappa$ is a threshold of function \eqref{internal2}.
From \eqref{internal2}, we can further deduce that
\begin{equation}
\label{internal1u}
\ibu - \ibx \in \kappa \partial \|\ibu \|_1,
\end{equation}
where $\partial \|\ibu \|_1$ denotes the sub-differential of the $l_1$-norm term which is equal to its gradient ($\pm 1$) at the differentiable points and equal to a set $[-1,1]$ at zero. Thus, LCA defines its dynamics on $\ibu$ as
\begin{equation}
\frac{d \ibu}{dt}= - \ibu + \ibx + \bPhi^\mathrm{T}(\ibb - \bPhi \ibx).
\label{eqn:dyna}
\end{equation}
It should be noticed that if the dynamics are defined as ${d \ibx}/{dt}$,
we need to implement $\partial \|\ibx \|_1$ which is equal to a set at the zero point. Therefore, LCA uses ${d \ibu}/{dt}$ rather than ${d \ibx}/{dt}$. The properties of LCA has been analysed in~\cite{rozell2008sparse,balavoine2012convergence,balavoine2011global}. Besides, LCA cannot directly handle the constrained problem in \eqref{eq-bp}.

\subsection{LPNN-LCA Framework}
LPNN is a framework for solving the general nonlinear constrained optimization problems. LCA can effectively handle the non-differentiable term in dynamics. But neither LPNN nor LCA can solve the problem in \eqref{eq-bp} alone. Hence, in our previous study, we have combined these two methods and devised the LPNN-LCA framework which can be used to solve the problem in~\eqref{eq-bp} \cite{feng2017lagrange}.

For this method, we first construct the Lagrangian of \eqref{eq-bp} as
\beq\label{eq-2}
L(\ibx,\blam)=\|\ibx\|_1+\blam^ \mathrm{ T }(\bPhi\ibx-\ibb),
\eeq
Then, follows the concept of LPNN in~\eqref{eq-1.8}, we define its dynamics:
\begin{subequations}\label{eq-dynamics0}
\beq
\frac{d\ibx}{dt}\!&=&\! -\frac{\partial L(\ibx,\blam)}{\partial \ibx}= -\partial\|\ibx\|_1-\bPhi^ \mathrm{ T }\blam, \\
\frac{d\blam}{dt}\!&=&\!\frac{\partial L(\ibx,\blam)}{\partial \blam}=\Phi\ibx-\ibb.
\eeq
\end{subequations}
To handle the non-differential part, the concept of LCA is utilized and its dynamics can be further modified as
\begin{subequations}\label{eq-dynamics1}
\beq
\frac{d\ibu}{dt}&=& -(\ibu-\ibx+\bPhi^ \mathrm{ T }\blam), \\
\frac{d\blam}{dt}&=&\Phi\ibx-\ibb.
\eeq
\end{subequations}
The relationship between $\ibx$ and $\ibu$ is given by~\eqref{internal2}. However, according to our preliminary simulation result, this system may not be stable.
Hence we further introduce an augmented term $\frac{1}{2}\|\bPhi\ibx-\ibb\|_2^2$ into its Lagrangian, then
\beq\label{eq-2_1}
L(\ibx,\blam)=\|\ibx\|_1+\blam^ \mathrm{ T }(\bPhi\ibx-\ibb)+\frac{1}{2}\|\bPhi\ibx-\ibb\|_2^2,
\eeq
The augmented term does not affect the objective value at an equilibrium point $\ibx^*$, because $\bPhi\ibx^*-\ibb=\mathbf{0}$. And it can further improve the stability of the system. Thus, the dynamics can be rewritten as
\begin{subequations}\label{eq-dynamics2}
\beq
\frac{d\ibu}{dt}&=& -(\ibu-\ibx+\bPhi^ \mathrm{ T }\blam)-\bPhi^ \mathrm{ T }(\bPhi\ibx-\ibb), \\
\frac{d\blam}{dt}&=&\Phi\ibx-\ibb.
\eeq
\end{subequations}
After the dynamics in \eqref{eq-dynamics2} settle down, the solution of problem \eqref{eq-bp} can be obtained at an equilibrium point. Even the augmented term is used, we can only prove the local asymptotic convergence of this system.
\subsection{Projection Neural Network}
The projection neural network \cite{xia2002projection,xia2004general} is based on the projection theorem~\cite{projectiontheorem}. It is devised to solve the following variational inequality $VI(\bU,\bZ)$, finding $\ibz^*\in\bZ$ such that
\beq
\bU(\ibz^*)^\mathrm{T}(\ibz-\ibz^*)\geq 0, \,\, for \,\,\forall\ibz\in\bZ.
\label{eq-VI}
\eeq
Where $\bU(\ibz):\mathbb{R}^p \to\mathbb{R}^p$ is a continuous function, the set $\bZ=\{\ibz\in\mathbb{R}^p|d_i\leq z_i\leq h_i, i=1,...,p\}$ is convex.
According to the optimization literature~\cite{xia2002projection}, problem~\eqref{eq-VI} is equivalent to the nonlinear projection formulation:
\beq
P_{\bZ}(\ibz-\bU(\ibz))=\ibz,
\label{eq-VI2}
\eeq
where $P_{\bZ}:\mathbb{R}^p \to \bZ$ is a projection operator defined by
\beq
P_{\bZ}(\ibz)= arg\min\limits_{\ibv \in \bZ} \|\ibz-\ibv\|, \nonumber
\eeq
$\|\cdot\|$ denotes $l_2$-norm.

The dynamic of projection neural network is defined as follows:
\beq
\frac{d\ibz}{dt}= \zeta\{P_{\bZ}(\ibz-\bU(\ibz))-\ibz\}
\label{eq-projection}
\eeq
where $\ibz\in\mathbb{R}^p$, and $\zeta$ is a positive scaling constant. This neural network can be used to solve the VI problem, by solving its relevant nonlinear projection formulation in \eqref{eq-VI2}. Projection neural network is a kind of recurrent neural network.

\section{Development of Proposed Algorithm} \label{section3}
\subsection{The Improved LPNN-LCA Framework}
In \cite{feng2017lagrange}, we have proved that the equilibrium point of the dynamics~\eqref{eq-dynamics2} is equivalent to the optimal solution of problem~\eqref{eq-bp}, and the system has local asymptotic convergence. However, its global convergence is hard to prove. In this section, we further modify the dynamics. After the modification, we hope that the equilibrium point of the new dynamics is still equivalent to the minimal solution of problem~\eqref{eq-bp} and the global convergence of the proposed neural network can be proved.

According to the description in Section~\ref{section2}, we see that the dynamic of the projection neural network is constructed by two parts: the projection operation term and a non-projected term.
It is worth noting that the operation $\ibu-\ibx$ in LPNN-LCA framework can also be seen as an projection operation. Because according to the threshold function in~\eqref{internal2}, if we let $\kappa=1$, the element-wise relationship between $\ibx$ and $\ibu$ is
\begin{equation}
u_i-x_i = \left\{ \begin{array}{lcl}
&u_i, & |u_i| \leq 1, \\
&1,  & u_i > 1,\\
&-1, & u_i < -1.
\end{array}\right.
\label{internal5}
\end{equation}
Apparently, the operation $u_i-x_i$ in~\eqref{internal5} can be seen as a projection operation $g_{\Gamma}(\cdot)$, which can project $u_i$ into a box set $[-1,1]$. Inspired by the dynamic of the projection neural network, we also modify the dynamics in~\eqref{eq-dynamics1} with the similar trick, after that the dynamics can be expressed as
\begin{subequations}\label{eq-dynamics3}
\beq
\frac{d\ibu}{dt}&=& -(\ibu-\ibx + \bPhi^\mathrm{T}\blam) = M(t), \\
\frac{d\blam}{dt}&=&\bPhi\ibx-\ibb-\bPhi(\ibu-\ibx + \bPhi^\mathrm{ T }\blam) \nonumber \\ &=&N(t)+\bPhi M(t).
\eeq
\end{subequations}
For the convenience of description, we respectively define that $M(t)=-(\ibu-\ibx + \bPhi^\mathrm{ T }\blam)$, $N(t)=\bPhi\ibx-\ibb$. While $N(t)$ is the original function given by (\ref{eq-dynamics1}b), it includes a non-projected term. Refer to the dynamic of the projection neural network in~\cite{xia2004general}, we also introduce a projection operation $\bPhi M(t)$ into it, thus we can get (\ref{eq-dynamics3}b). The matrix before the additional term is needed for resizing its dimensions. With the dynamics in \eqref{eq-dynamics3}, the parameters are updated with following steps:
\begin{subequations}\label{eq:update}
\beq
\ibu^{k+1}=\ibu^k+\mu\frac{d\ibu^k}{dt},\\
\blam^{k+1}=\blam^k+\mu\frac{d\blam^k}{dt},
\eeq
\end{subequations}
where $\mu$ is the step size ($\mu>0$).

\subsection{Properties Analysis}
The equilibrium points of the dynamics in (\ref{eq-dynamics3}) are with respect to the internal state variable $\ibu$, while the optimal solution of original problem~\eqref{eq-bp} is in regard to the decision variable $\ibx$. Hence, we need Theorem $1$ to show that the equilibrium point of the dynamics in (\ref{eq-dynamics3}) is identical to the optimal solution of problem~\eqref{eq-bp}.

\textbf{Theorem 1}:
{\it
Let $(\ibu^{*},\blam^{*})$ denotes an equilibrium point of~\eqref{eq-dynamics3}. At the equilibrium point, the KKT conditions of~\eqref{eq-bp} are satisfied. Since the optimization problem given in~\eqref{eq-bp} is convex, and the regularity condition is satisfied. Thus the equilibrium point of (\ref{eq-dynamics3}) is equivalent to the optimal solution of the problem in~\eqref{eq-bp}.}

Proof: The KKT conditions of problem \eqref{eq-bp} are
\begin{subequations}\label{eq-3}
\beq
\mathbf{0}&\in&\partial ||\ibx||_1 + \bPhi^\mathrm{ T } \blam,\\
\mathbf{0}&=&\bPhi\ibx-\ibb.
\eeq
\end{subequations}

According to the definition of the equilibrium point, we have
\begin{subequations}\label{eq-6.1}
\beq
M^{*}(t)=0 \\
\bPhi M^{*}(t)+N^{*}(t)=0
\eeq
\end{subequations}
Where $M^{*}(t)=-(\ibu^{*}-\ibx^{*} + \bPhi^\mathrm{ T }\blam^{*})$, $N^{*}(t)=\bPhi\ibx^{*}-\ibb$.
By the concept of LCA, we know that $\ibu^{*}-\ibx^{*}\in \partial ||\ibx^{*}||_1$, thus from (\ref{eq-6.1}a) we can deduce that：
\beq
\mathbf{0} \in \partial ||\ibx^{*}||_1 + \bPhi^\mathrm{ T }\blam^{*}.
\label{eq-equilibrium1.2}
\eeq
That means (\ref{eq-3}a) is satisfied at the equilibrium point. Then from \eqref{eq-6.1}, it is obvious that
\beq
N^{*}(t)=\bPhi\ibx^{*}-\ibb=\mathbf{0},
\label{eq-equilibrium2}
\eeq
(\ref{eq-3}b) is satisfied. So the equilibrium point of~\eqref{eq-dynamics3} satisfies the KKT conditions of problem~\eqref{eq-bp}. Besides, we know the problem in~\eqref{eq-bp} is convex and the regularity condition is satisfied. Hence, the KKT conditions in \eqref{eq-3} are sufficient and necessary. Moreover, we can say that the equilibrium point of~\eqref{eq-dynamics3} is a global optimal solution of problem~\eqref{eq-bp}. $\blacksquare$

According to the analysis of LPNN given in \cite{zhang1992lagrange}, we know that (\ref{eq-dynamics1}a) is used for seeking the minimum objective value, while (\ref{eq-dynamics1}b) aims at making the variables fall into the feasible region. After we modify the dynamics, these features are still valid. However, comparing with dynamics in \eqref{eq-dynamics1}, (\ref{eq-dynamics3}) accelerates its convergence to the optimal point, and it does not introduce any augmented term. After modification the complexity of the dynamics is still equal to ${\cal O}(mn)$. We know that the circuit complexity of analog neural network depends on the time derivative calculations. Hence, the proposed improved LPNN-LCA framework basically does not increase the circuit complexity.

\section{Global Stability Analysis} \label{section4}
According to Theorem 1, when the dynamics in \eqref{eq-dynamics3} settle down, the equilibrium point is an optimal solution of problem \eqref{eq-bp}. Next, we prove that the proposed neural network has global stability.

\textbf{Lemma 1}:
{\it
$\Gamma\subset\mathbb{R}^n$ is a closed and convex set, $g_{\Gamma}(\cdot)$ denotes the projection operator to $\Gamma$. For any $\ibv \in \mathbb{R}^n$ and $\ibv'\in\Gamma$,
\beq\label{eq-7}
(\ibv-g_{\Gamma}(\ibv))^\mathrm{T}(g_{\Gamma}(\ibv)-\ibv')\geqslant 0.
\eeq}

The proof is shown in \cite{projectiontheorem}.

For the proposed neural network, $g_{\Gamma}(\ibu)=\ibu-\ibx$ is a projection operation to a closed and convex set. Thus, if we let $\ibv=\ibu$ and $\ibv'=\ibu^*-\ibx^*$, the inequality~\eqref{eq-7} can be rewritten as
\beq\label{eq-8}
\ibx^\mathrm{T}[(\ibu-\ibx)-(\ibu^*-\ibx^*)]\geqslant 0.
\eeq
Then let $\ibv=\ibu^*$ and $\ibv'=\ibu-\ibx$, based on the inequality in~\eqref{eq-7}, we can deduce that
\beq\label{eq-9}
-\ibx^{*^\mathrm{T}}[(\ibu-\ibx)-(\ibu^*-\ibx^*)]\geqslant 0.
\eeq
Add~\eqref{eq-8} and~\eqref{eq-9}, we have
\beq \label{eq-10}
(\ibx-\ibx^*)^\mathrm{T}[(\ibu-\ibx)-(\ibu^*-\ibx^*)]\geqslant 0
\eeq

\textbf{Lyapunov stability theory}:
{\it
Let $\ibv^*$ be an equilibrium point of the system $\dot{\ibv} = \hat{f}(\ibv)$, Let $V:\mathbb{R}^n \to \mathbb{R}$ be a continuously differentiable function such that
\beq
&&V(\mathbf{0})=0 \,\,\,\, and \,\,\,\, V(\ibv)\geq 0, \,\,\,\, \forall \ibv \neq \mathbf{0} \nonumber \\
&&\|\ibv\|\to\infty \,\,\,\,\Rightarrow \,\,\,\,V(\ibv) \to \infty \nonumber \\
&&\dot{V}(\ibv)< 0, \,\,\,\, \forall \ibv \neq \mathbf{0} \nonumber
\eeq
then the system is globally asymptotically stable in the sense of Lyapunov.}

With Lemma 1 and the Lyapunov stability theory, we can deduce the following theorem.

\textbf{Theorem 2}:
{\it
The analog neural network given in \eqref{eq-dynamics3} is globally asymptotically stable in the sense of Lyapunov and it is globally convergent to a solution of problem \eqref{eq-bp}.}

Proof:
To prove the global convergence, first, we need to devise a valid Lyapunov function. Here we use the standard form $V:\mathbb{R}^n \to \mathbb{R}$ given by
\beq\label{eq-11}
V(\ibw-\ibw^{*})=\frac{1}{2}||\ibw-\ibw^{*}||_{2}^{2}
\eeq
where $\ibw=(\ibu,\blam)^\mathrm{T}$, and $\ibw^{*}=(\ibu^*,\blam^*)^\mathrm{T}$ is an equilibrium point of dynamics in \eqref{eq-dynamics3}.
According to Theorem 1, we know that $\ibw^{*}$ is also the optimal solution of problem \eqref{eq-bp}.
Obviously, the function $V(\ibw-\ibw^{*})$ is continuous, differentiable and radially unbounded. When $\ibw=\ibw^*$, we have $V(\mathbf{0})=0$. Otherwise, $V(\ibw-\ibw^{*})\geq 0$, for any $\ibw \neq \ibw^*$.
Next, we prove $\dot{V}(\ibw-\ibw^{*})<0$, for any $\ibw \neq \ibw^*$.
The derivative of $V(\ibw-\ibw^{*})$ is given by
\beq\label{eq-12}
&&\dot{V}(\ibw-\ibw^{*})=\frac{d\,V(\ibw-\ibw^*)}{d\,t} \nonumber\\
&&=(\ibw-\ibw^{*})^\mathrm{T}\frac{d\,\ibw}{d\,t} \nonumber\\
&&=
\left(
\begin{matrix}
\ibu-\ibu^*\\
\blam-\blam^*
\end{matrix}
\right)^\mathrm{T}
\left(
\begin{matrix}
M(t)\\
\bPhi M(t)+N(t)
\end{matrix}
\right)\nonumber\\
&&=(\ibu-\ibu^*)^\mathrm{T}M(t)+(\blam-\blam^*)^\mathrm{T}\left(\bPhi M(t)+N(t)\right)\nonumber\\
&&=-M(t)^\mathrm{T}M(t)+(\ibx-\ibx^*)^\mathrm{T}M(t)+(\blam-\blam^*)^\mathrm{T}N(t)\nonumber \\
&&=-M(t)^\mathrm{T}M(t)-(\ibx-\ibx^*)^\mathrm{T}[(\ibu-\ibx)-(\ibu^*-\ibx^*)]
\eeq

Apparently, from \eqref{eq-10} we know that, for any $\ibw \neq \ibw^*$, the function in~\eqref{eq-12} are less than $0$.
So $\dot{V}(\ibw-\ibw^*)<0$ for any $\ibw\neq\ibw^*$.
Thus we can say that the proposed neural network is globally asymptotically stable in the sense of Lyapunov. If $\dot{V}(\ibw-\ibw^*)=0$, we have $\ibw=\ibw^*$. Similar with the proof in \cite{liu2016l1}, we see the system in \eqref{eq-dynamics3} is globally convergent to an equilibrium point. $\blacksquare$

Similar with the dynamics in \eqref{eq-dynamics2}, we can also introduce an augment term $\frac{1}{2}\|\bPhi\ibx-\ibb\|_2^2$ into its Lagrangian, then we have
\begin{subequations}\label{eq-dynamics4}
\beq
\frac{d\ibu}{dt}&=&M(t)-\bPhi^\mathrm{T}N(t), \\
\frac{d\blam}{dt}&=&N(t)+\bPhi M(t).
\eeq
\end{subequations}
Compared with the dynamics in \eqref{eq-dynamics3}, \eqref{eq-dynamics4} further improve its convexity and  convergence speed.

\begin{figure*}[!ht]
\centering
\begin{tabular}{c@{\extracolsep{2mm}}c@{\extracolsep{2mm}}c@{\extracolsep{2mm}}c}
\mbox{\epsfig{figure=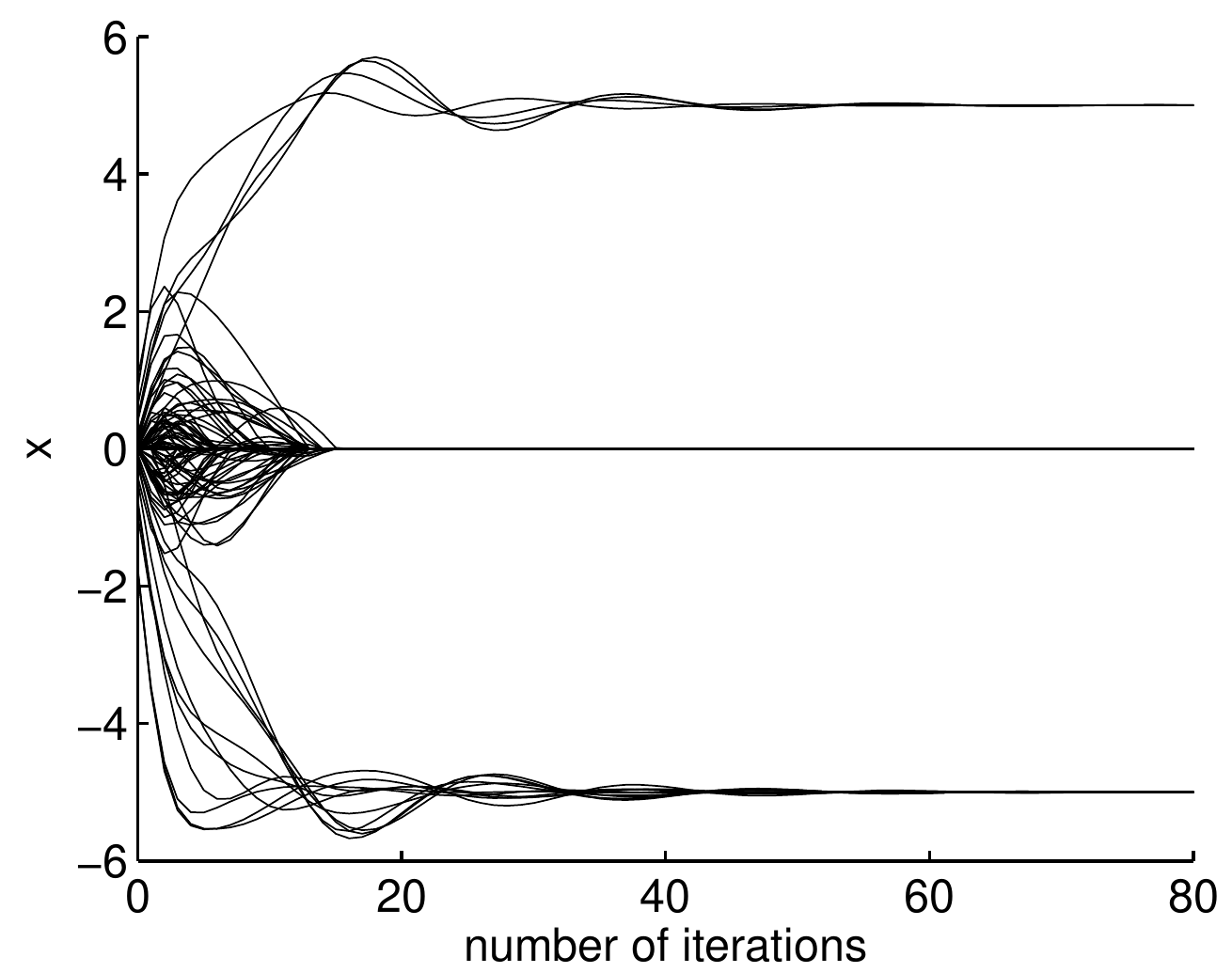,width=1.65in}} &
\mbox{\epsfig{figure=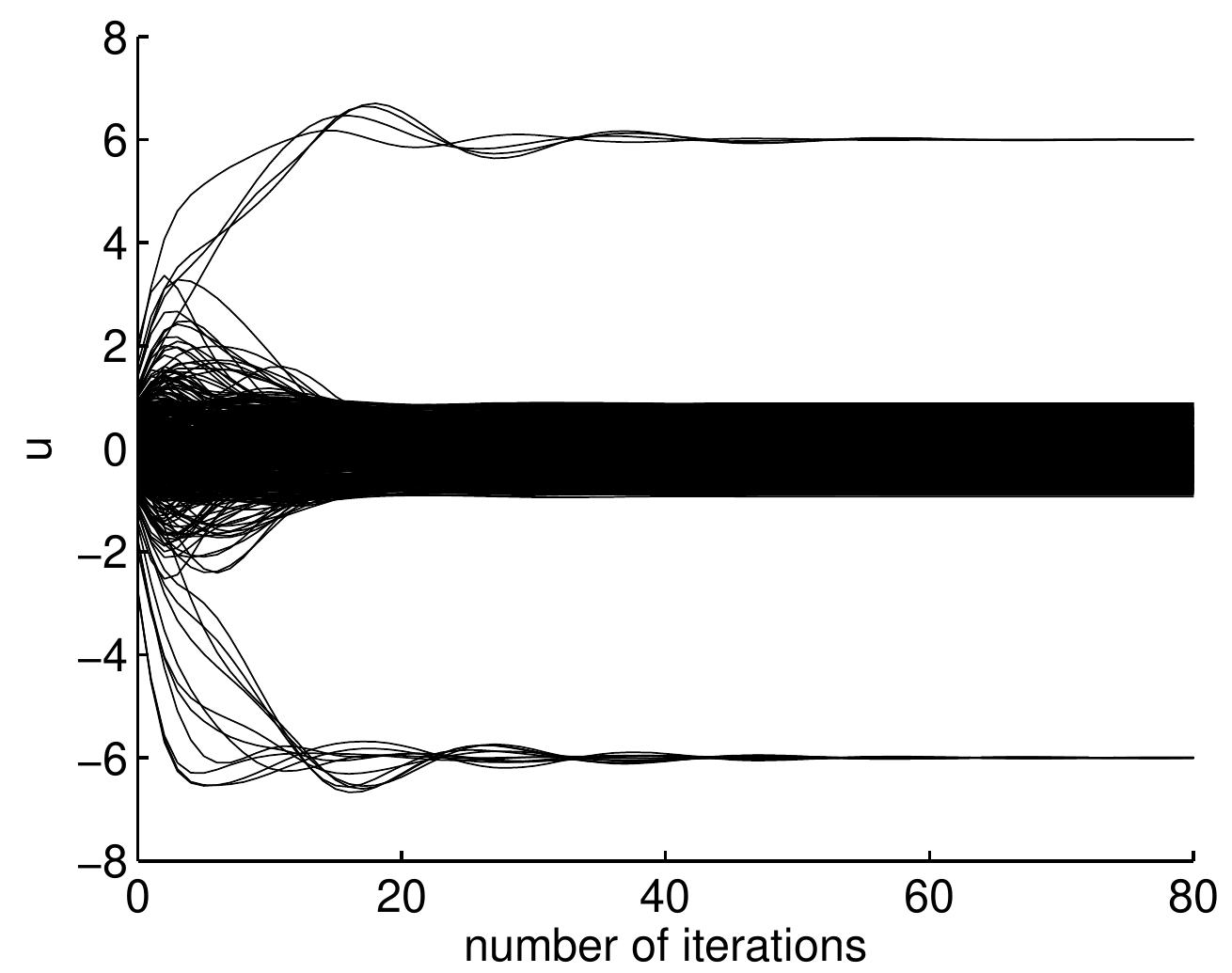,width=1.65in}} &
\mbox{\epsfig{figure=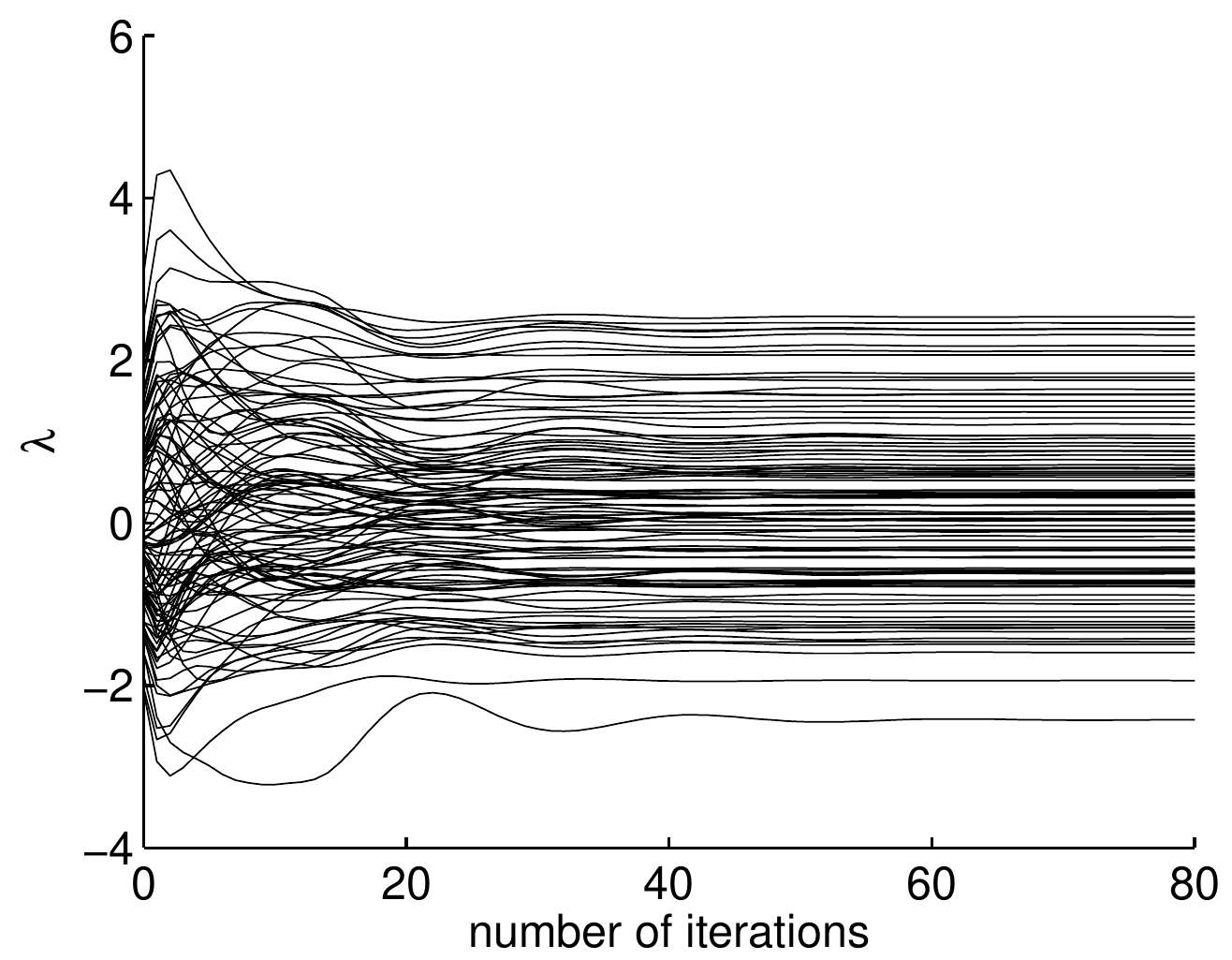,width=1.65in}} &
\mbox{\epsfig{figure=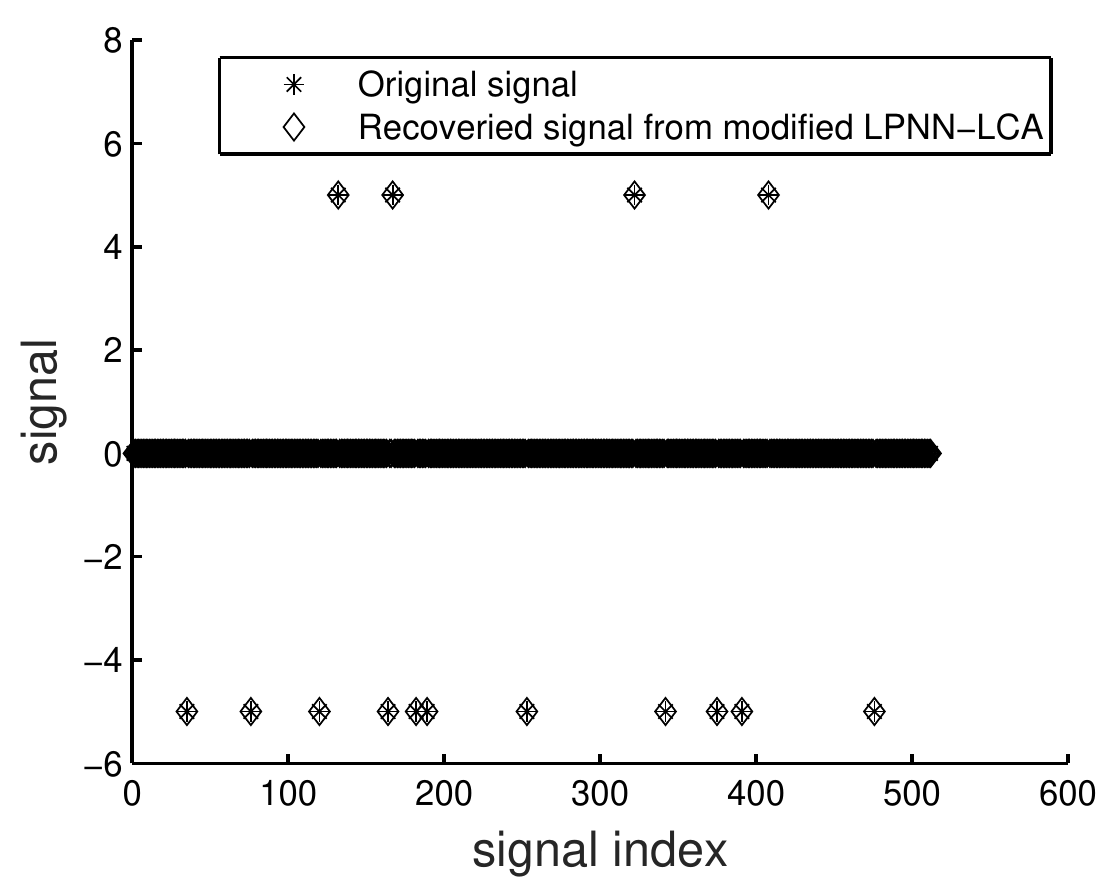,width=1.65in}}\\
\mbox{\epsfig{figure=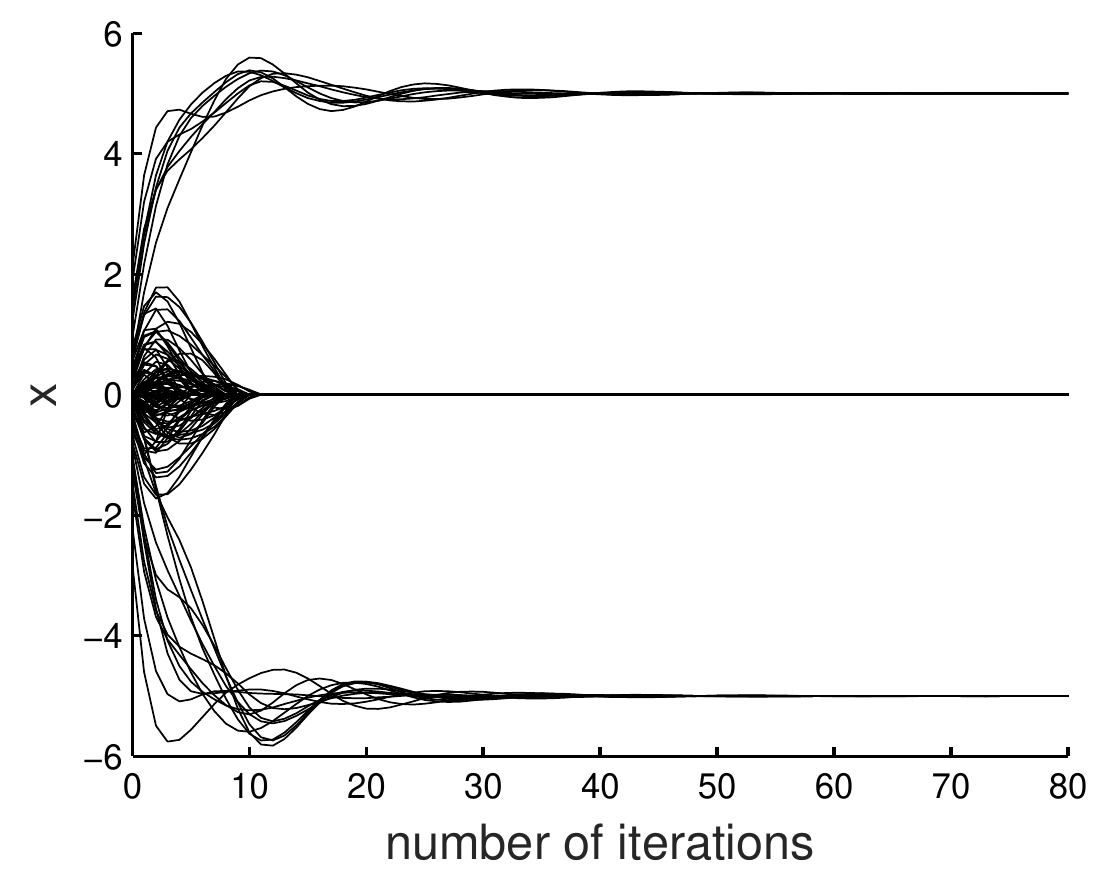,width=1.65in}} &
\mbox{\epsfig{figure=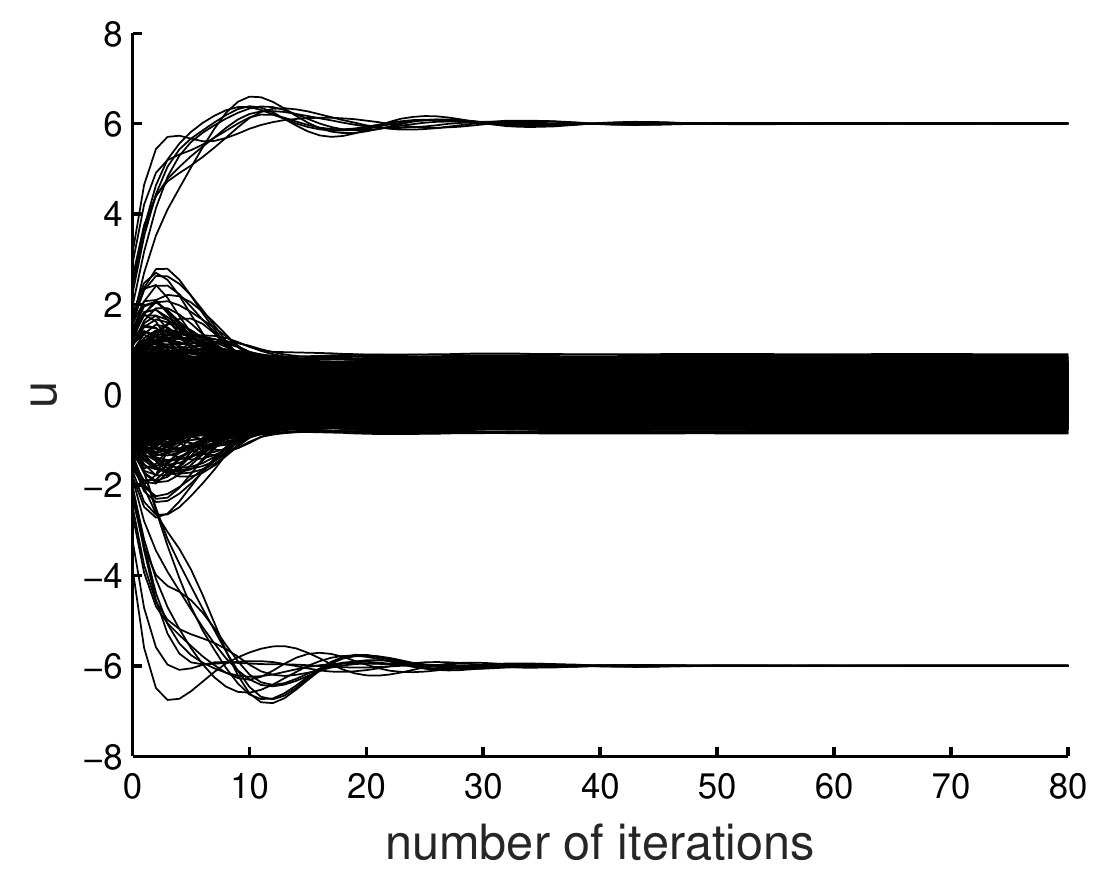,width=1.65in}} &
\mbox{\epsfig{figure=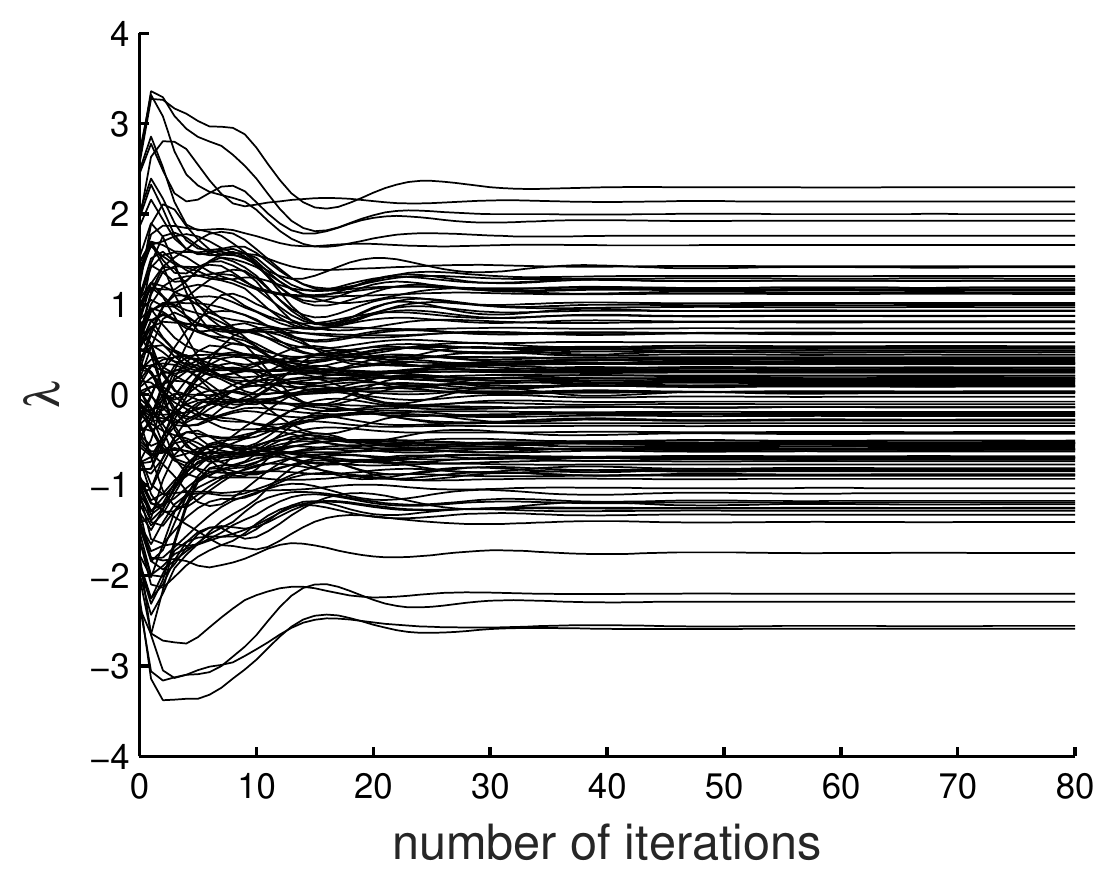,width=1.65in}} &
\mbox{\epsfig{figure=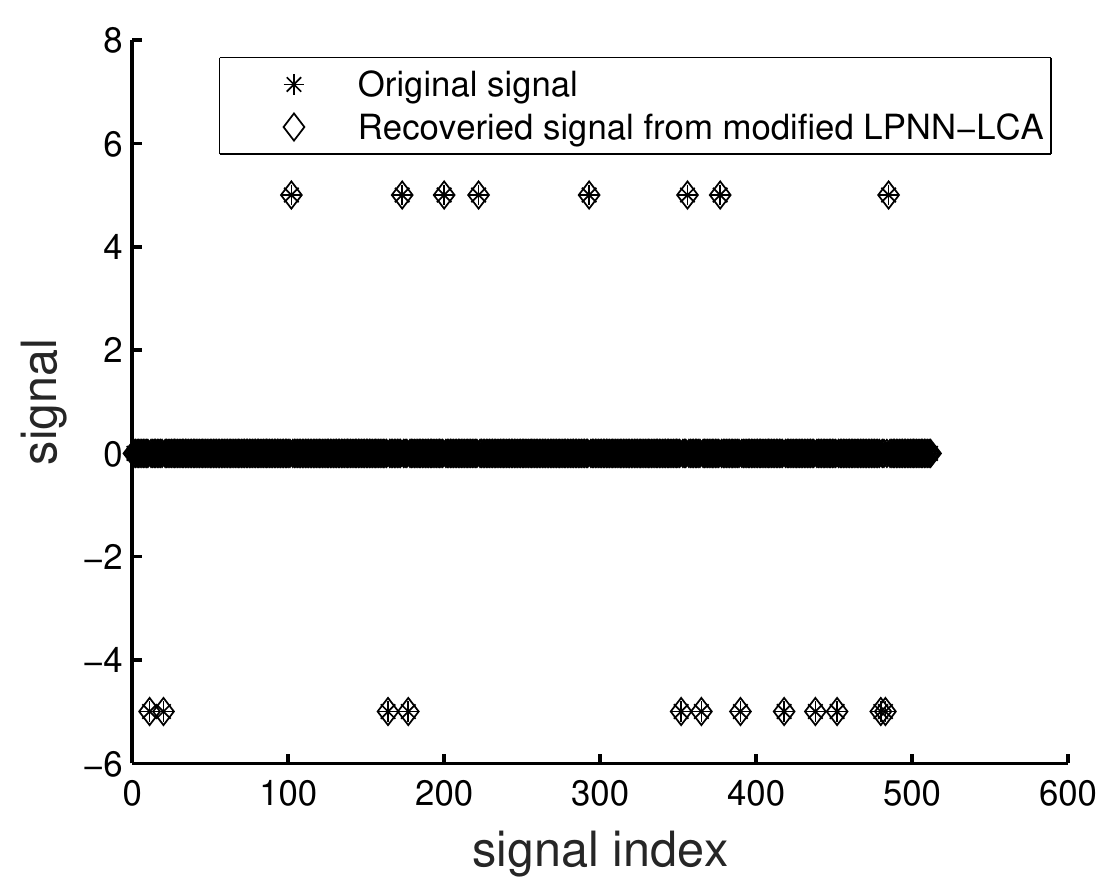,width=1.65in}}\\
\mbox{\epsfig{figure=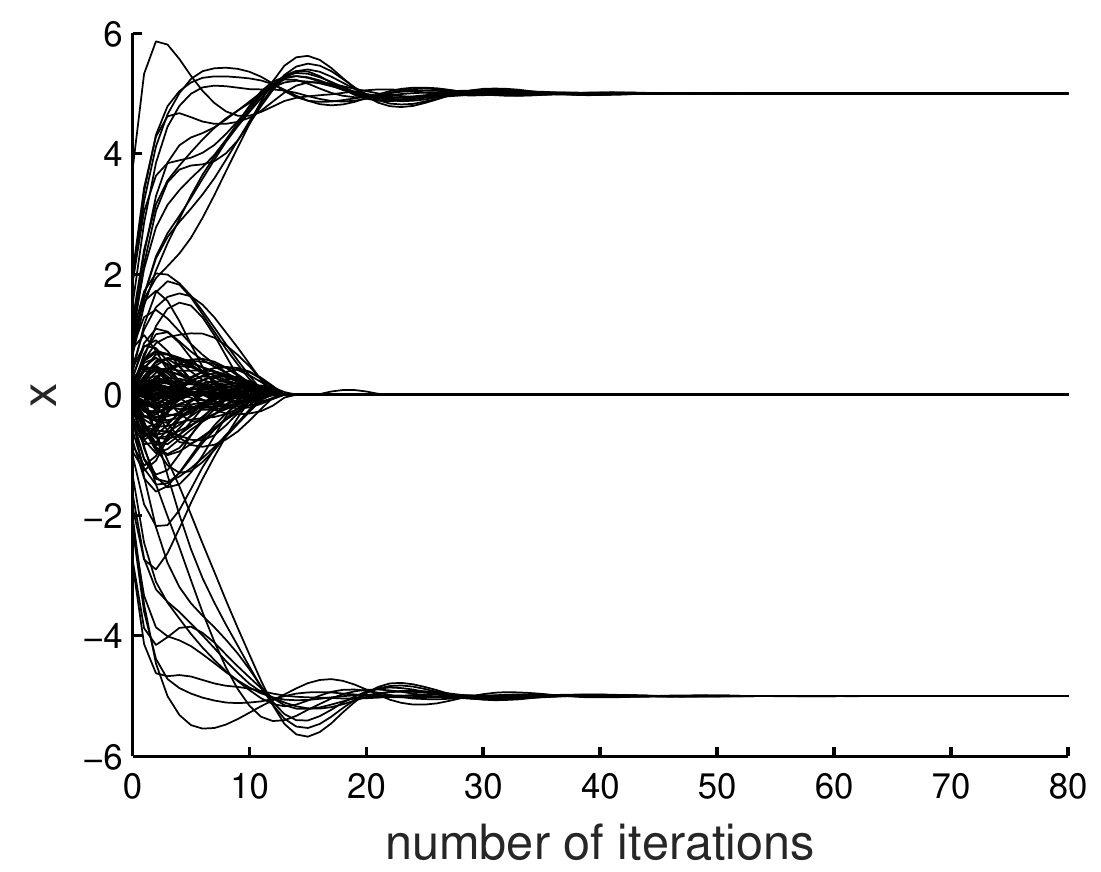,width=1.65in}} &
\mbox{\epsfig{figure=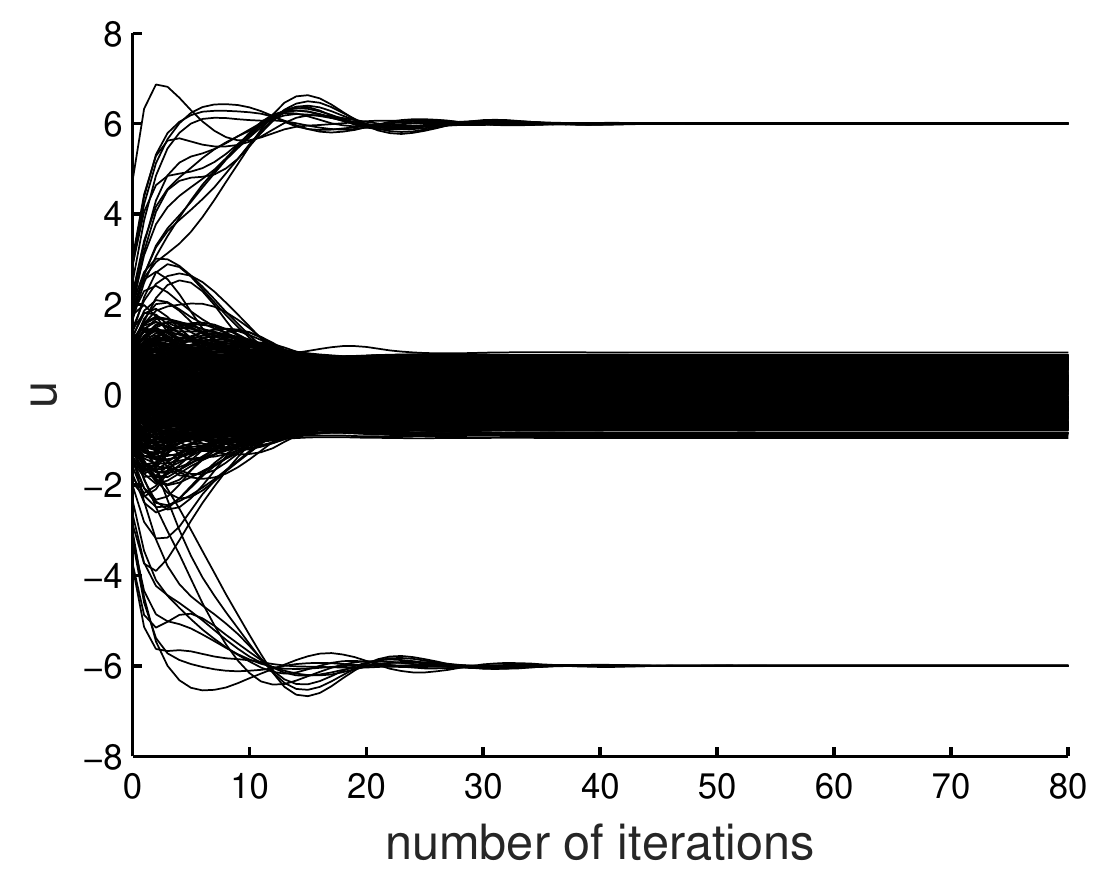,width=1.65in}} &
\mbox{\epsfig{figure=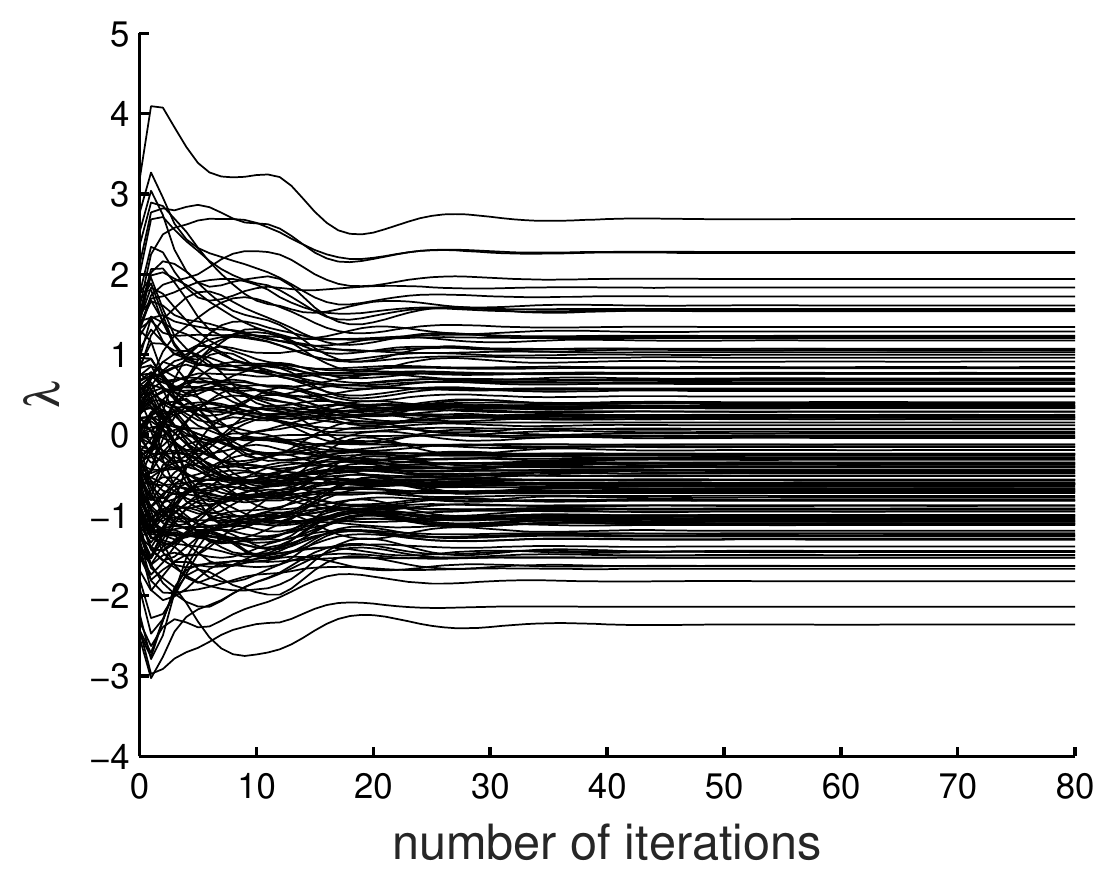,width=1.65in}} &
\mbox{\epsfig{figure=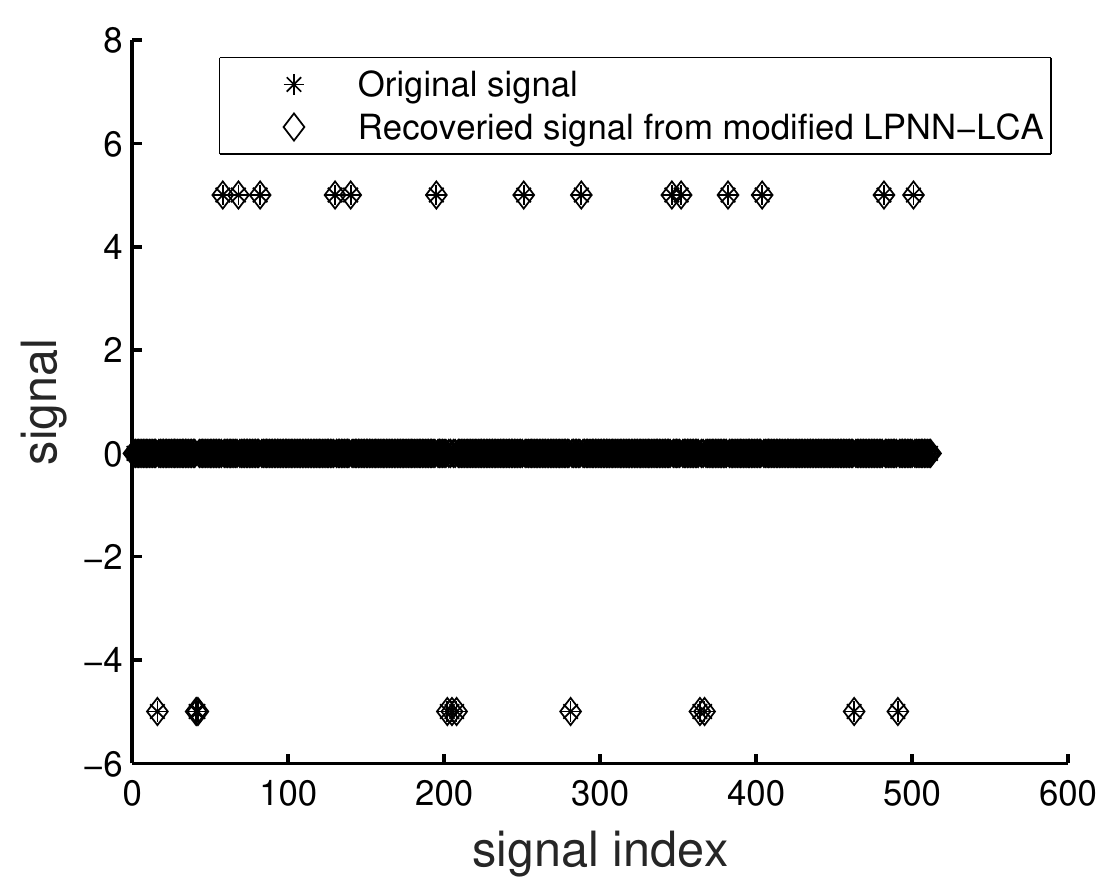,width=1.65in}}\\
\end{tabular}
\caption{Convergence of the proposed method when $n=512$. For the first row $m=100$ and $\Omega=15$. For the second row $m=130$ and $\Omega=20$. For the third row $m=160$ and $\Omega=25$.}
\label{fig:convergence512}
\end{figure*}

\begin{figure*}[!ht]
\centering
\begin{tabular}{c@{\extracolsep{2mm}}c@{\extracolsep{2mm}}c@{\extracolsep{2mm}}c}
\mbox{\epsfig{figure=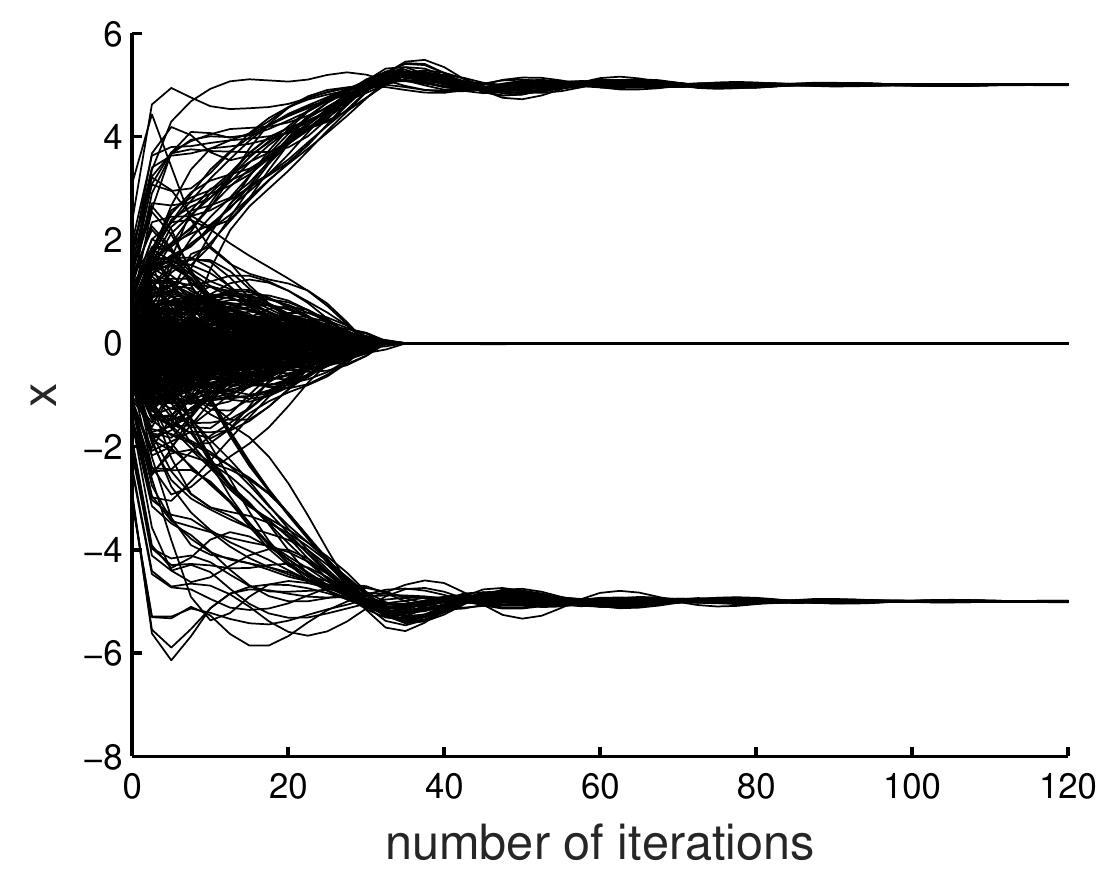,width=1.65in}} &
\mbox{\epsfig{figure=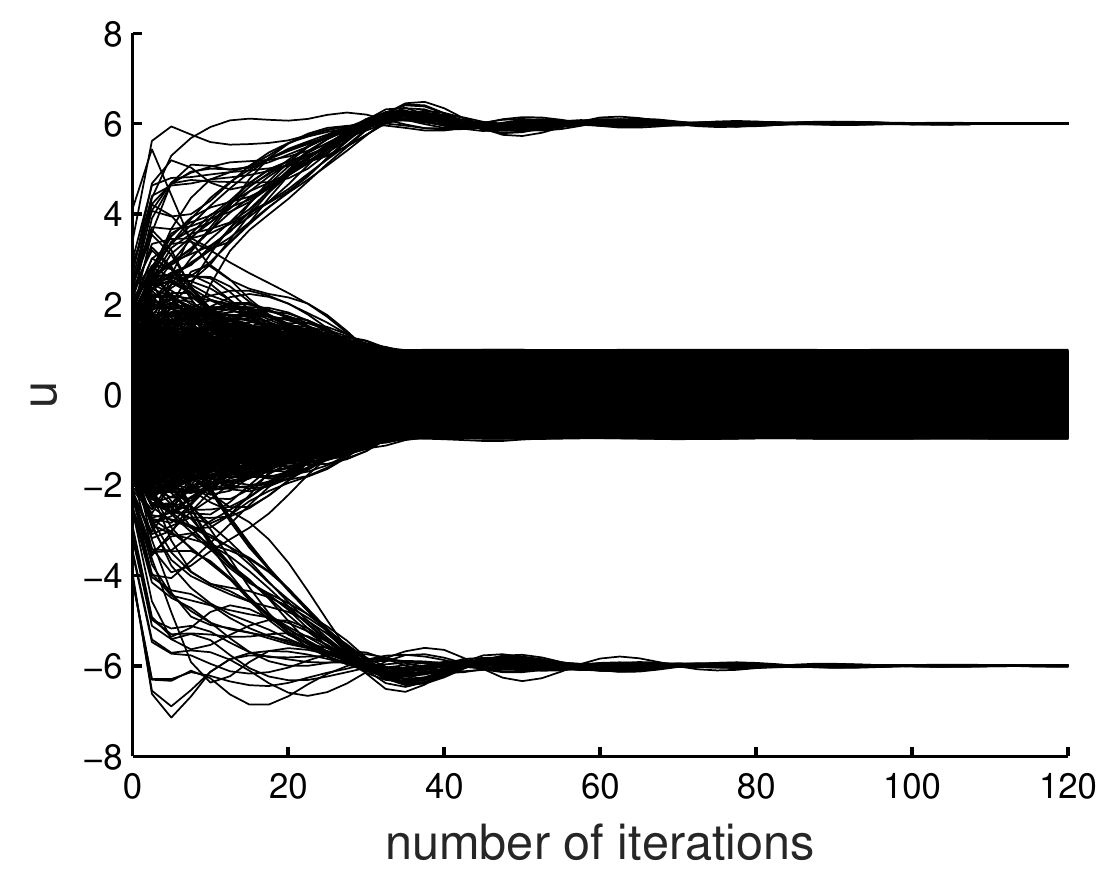,width=1.65in}} &
\mbox{\epsfig{figure=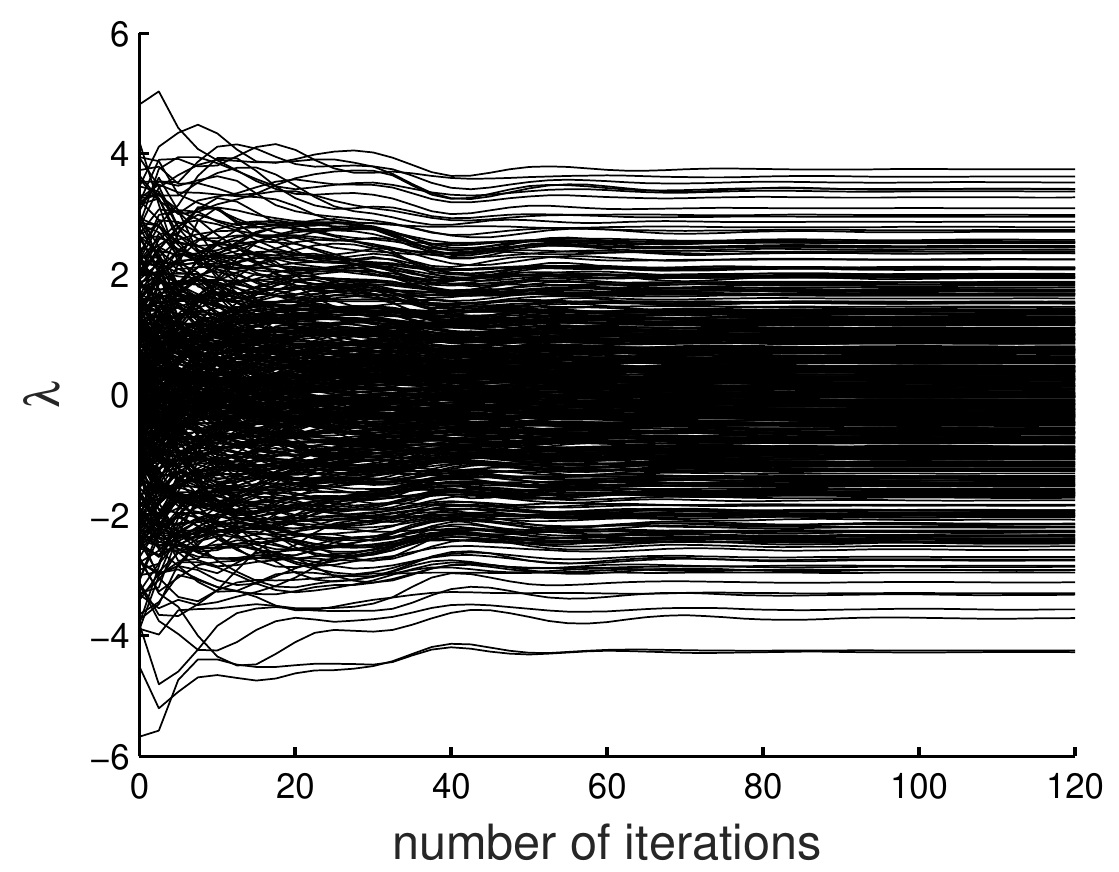,width=1.65in}} &
\mbox{\epsfig{figure=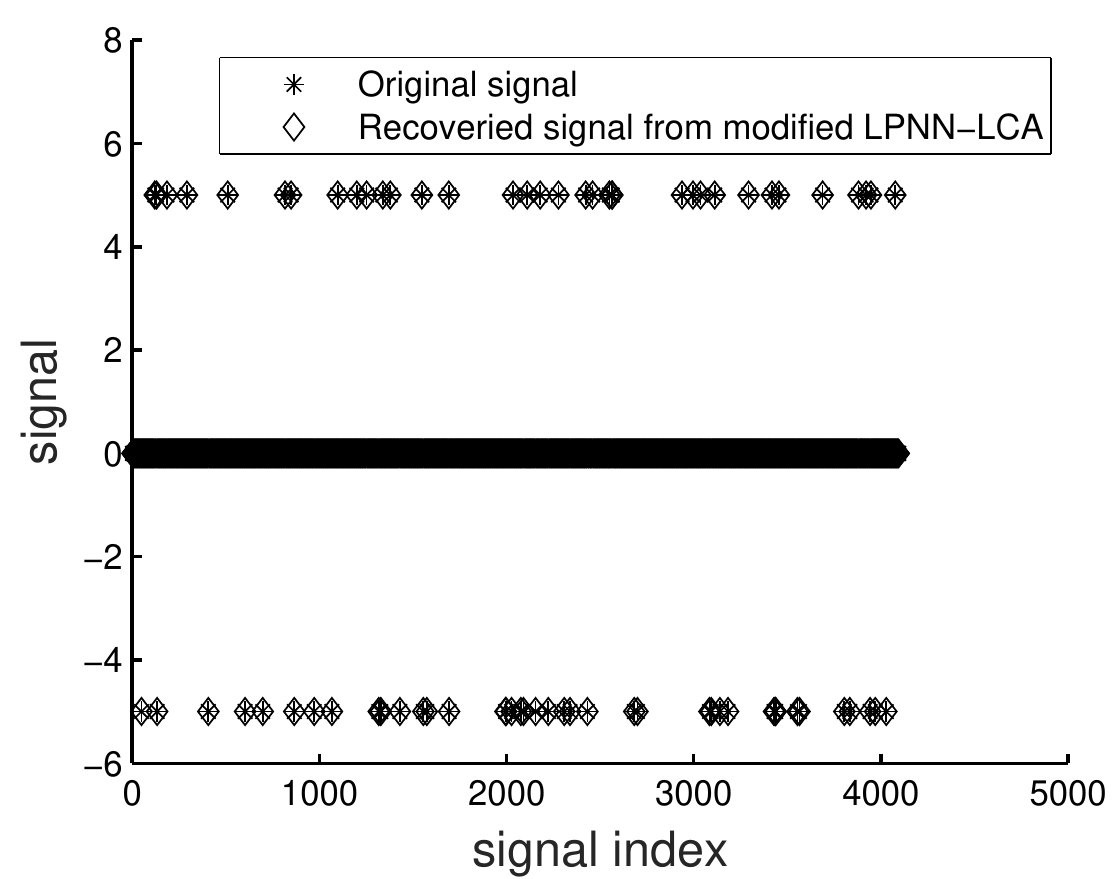,width=1.65in}}\\
\mbox{\epsfig{figure=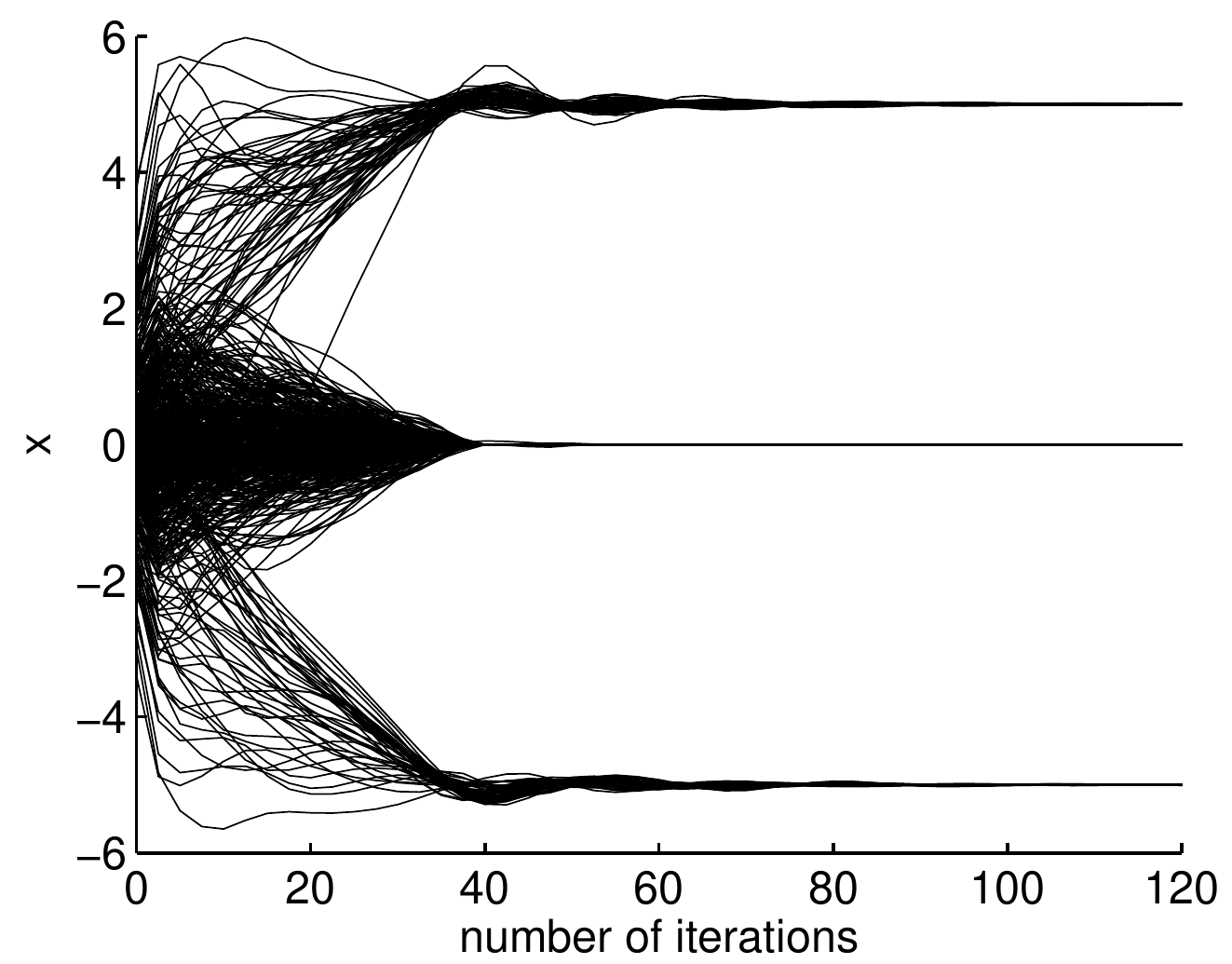,width=1.65in}} &
\mbox{\epsfig{figure=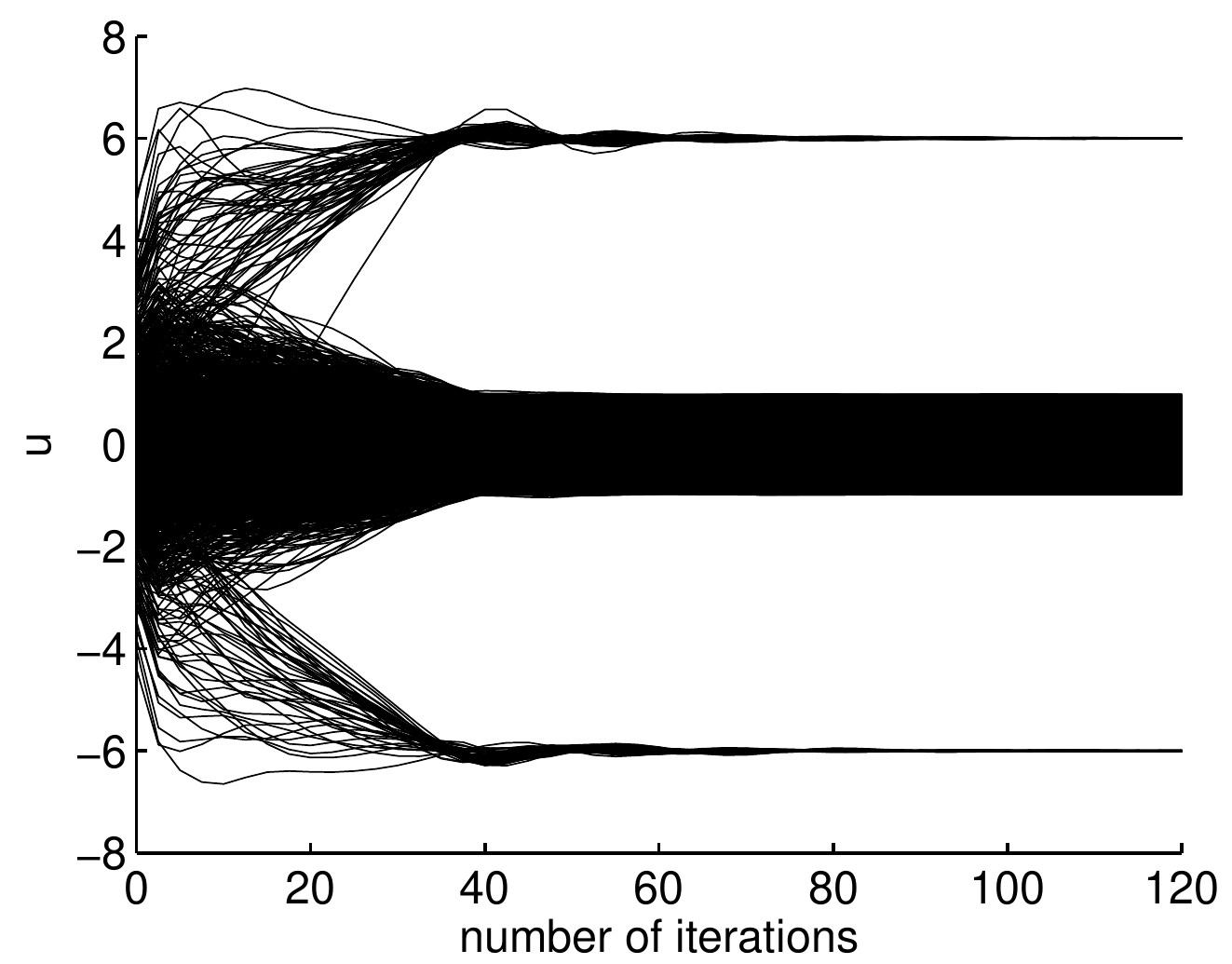,width=1.65in}} &
\mbox{\epsfig{figure=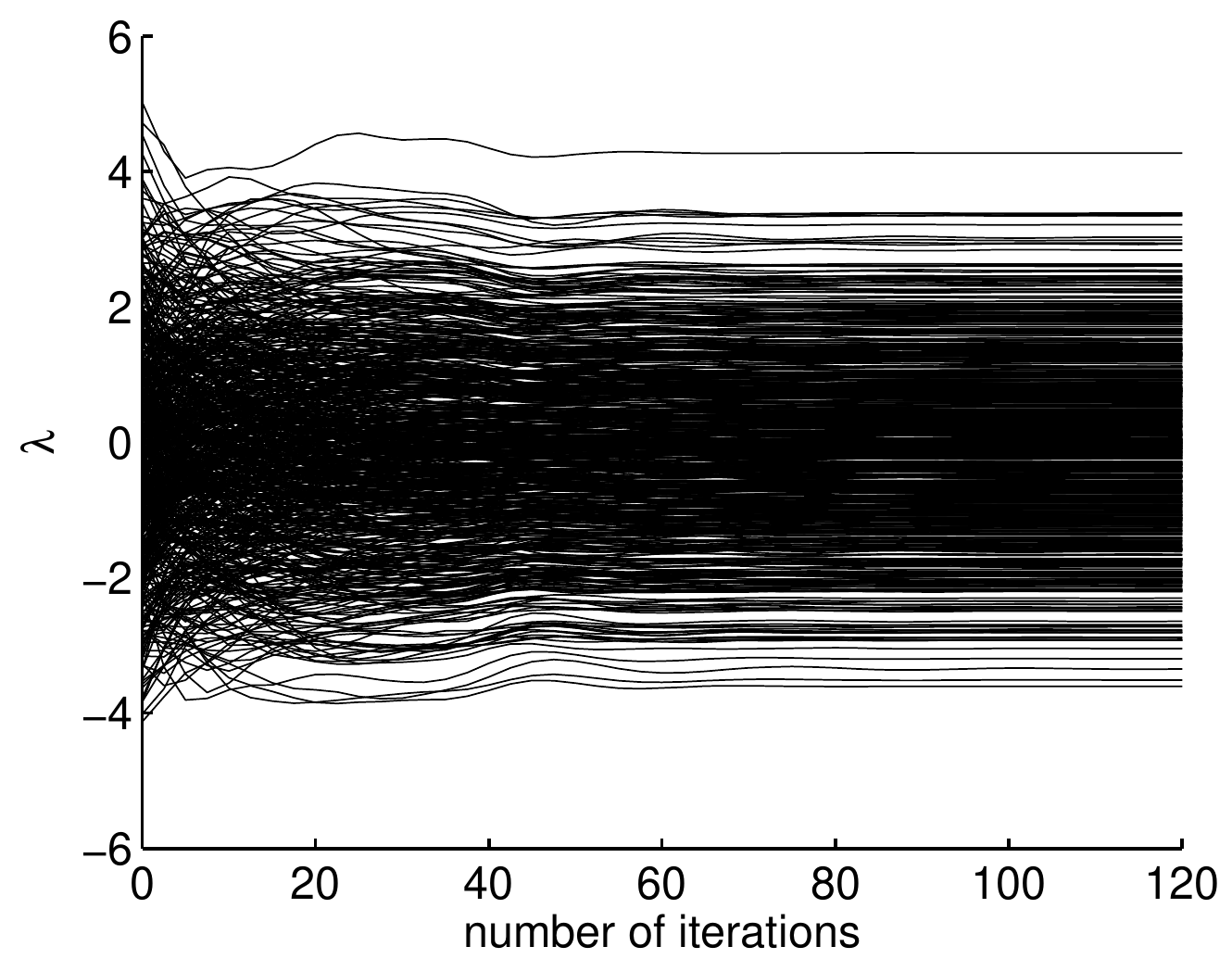,width=1.65in}} &
\mbox{\epsfig{figure=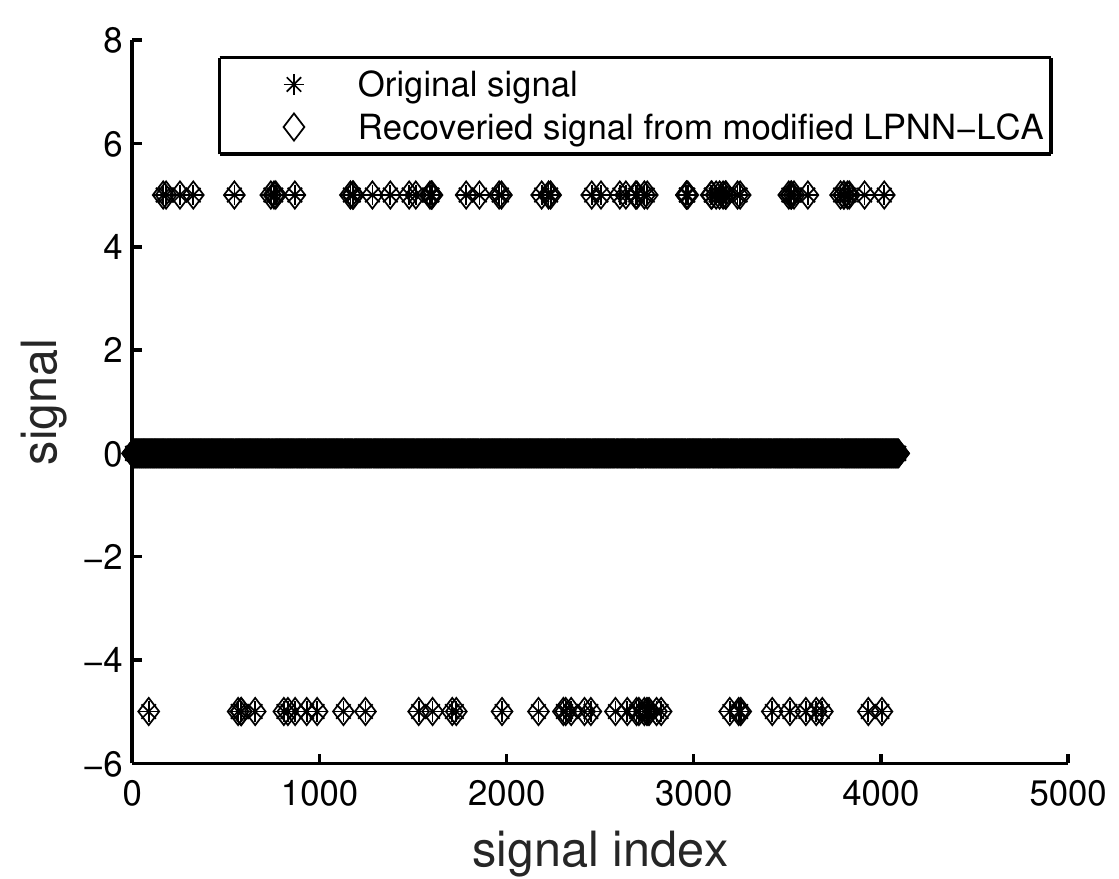,width=1.65in}}\\
\mbox{\epsfig{figure=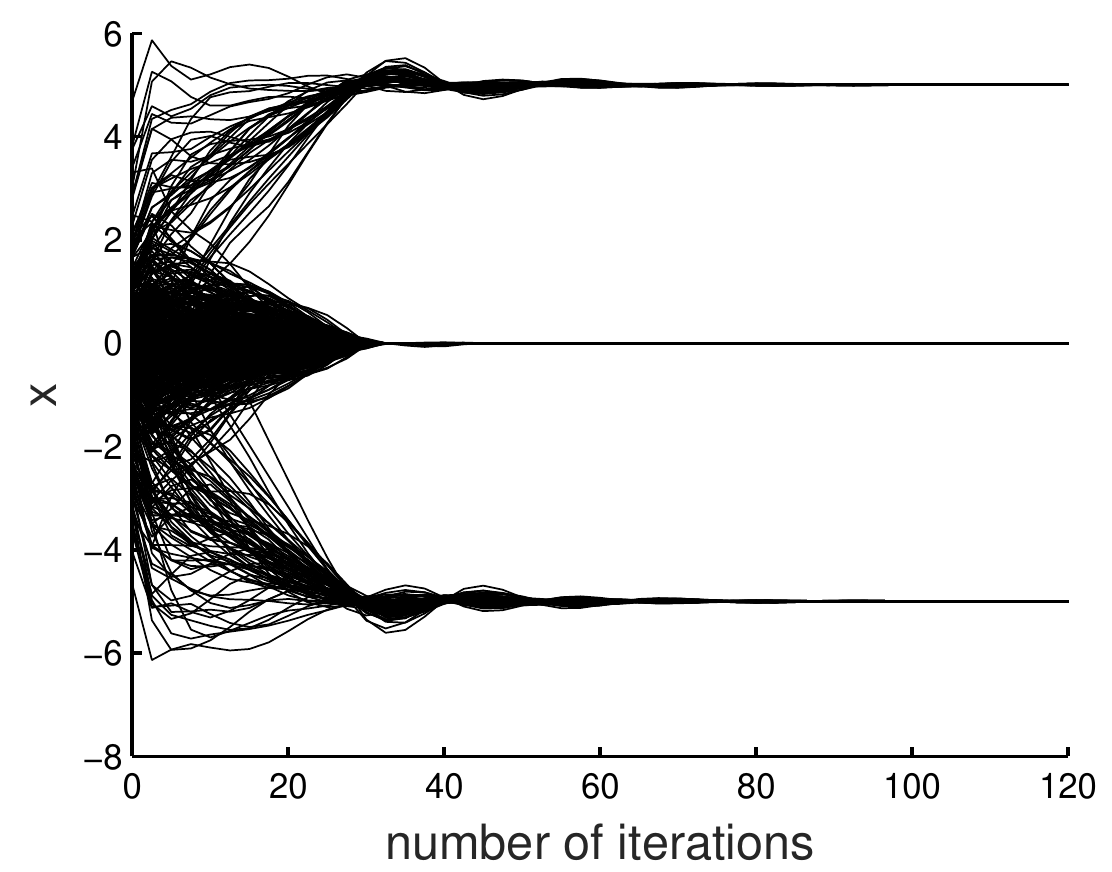,width=1.65in}} &
\mbox{\epsfig{figure=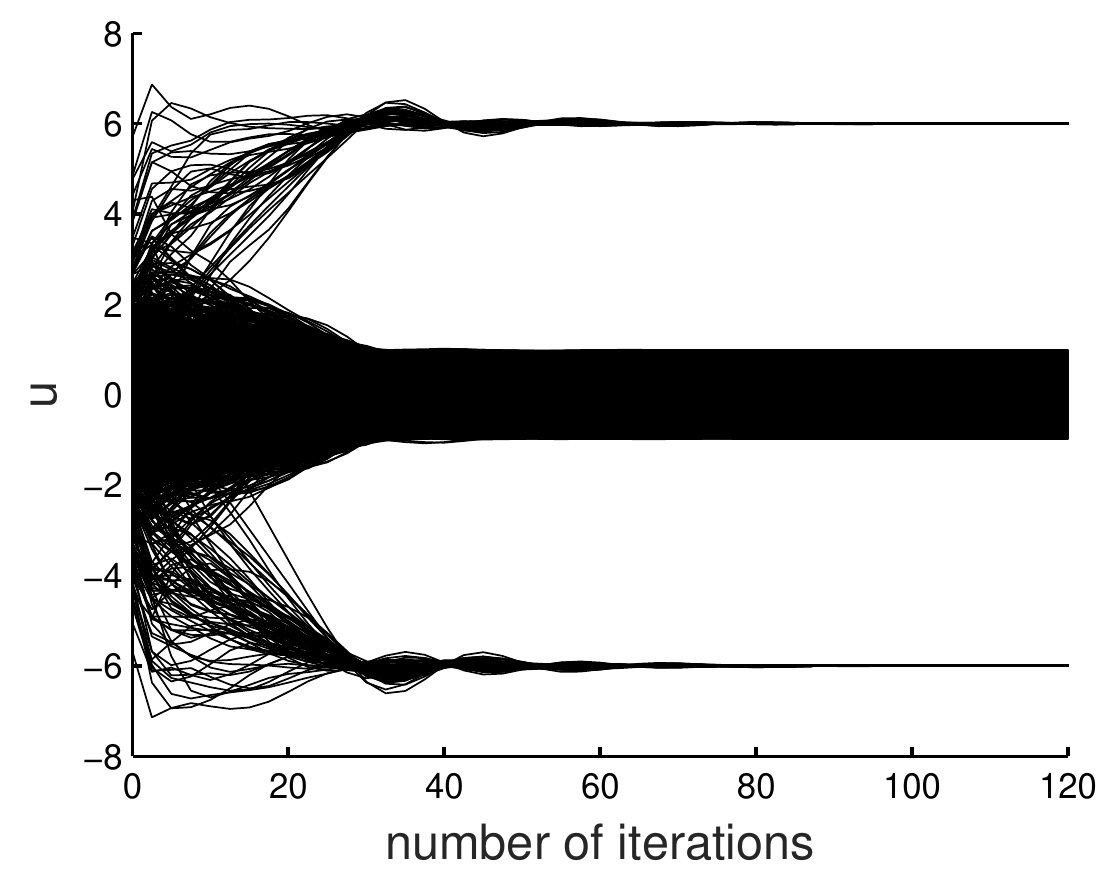,width=1.65in}} &
\mbox{\epsfig{figure=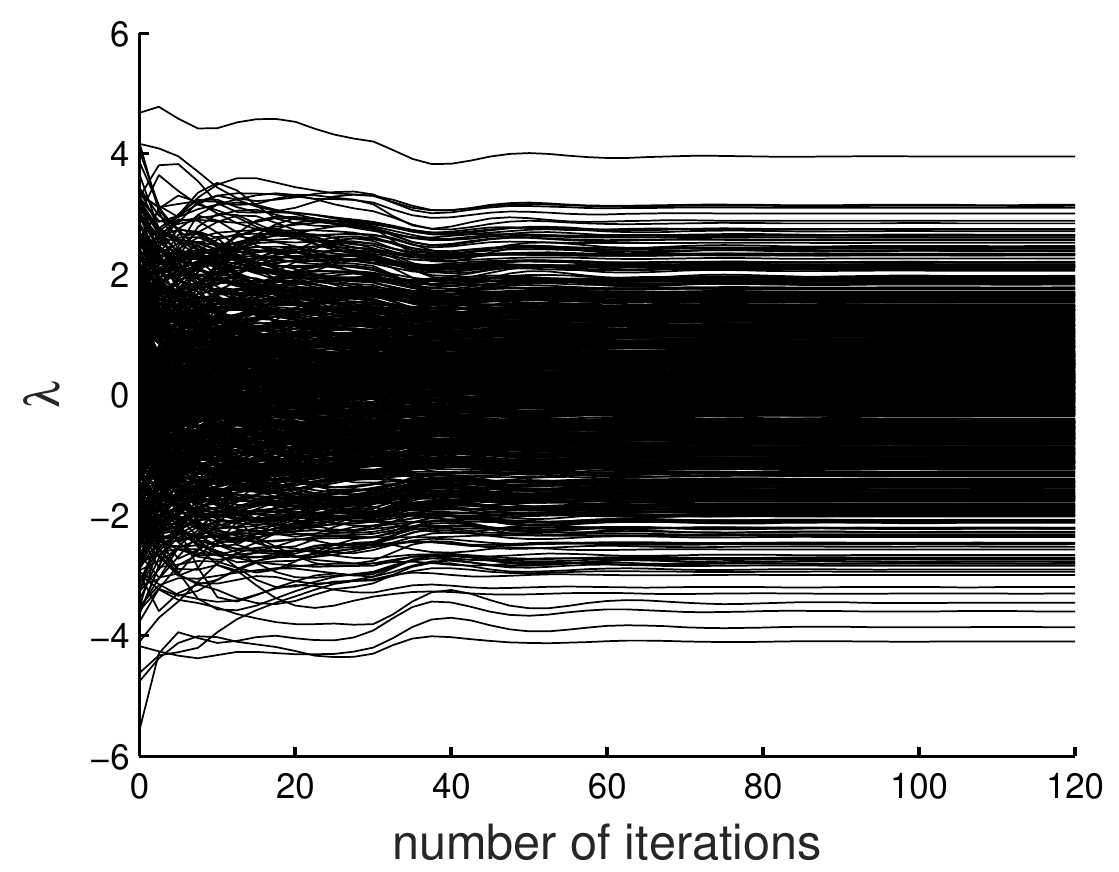,width=1.65in}} &
\mbox{\epsfig{figure=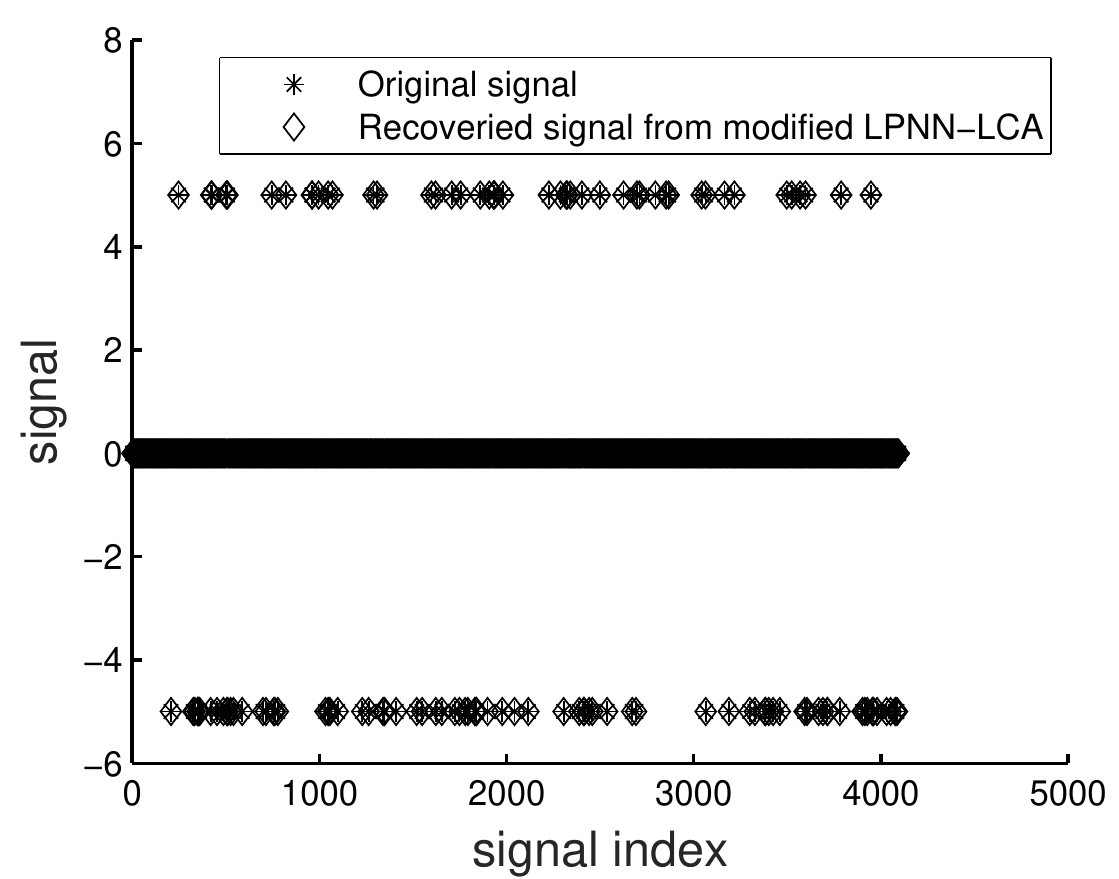,width=1.65in}}\\
\end{tabular}
\caption{Convergence of the proposed method when $n=4096$. For the first row $m=500$ and $\Omega=75$. For the second row $m=600$ and $\Omega=100$. For the third row $m=700$ and $\Omega=125$.}
\label{fig:convergence4096}
\end{figure*}

%

\section{Simulation and Experimental Result}\label{section5}
In this section, we conduct several experiments to evaluate the performance of the proposed algorithm. In the following experiments, the length of the signal $\ibx$ are $512$ or $4096$. When $n=512$, the number of non-zero elements $\Omega\in\{15,20,25\}$. And their position are randomly chosen with uniform distribution. Their corresponding values are randomly equal to $+5$ or $-5$. While for the case $n=4096$, the number of non-zero elements $\Omega\in \{75,100,125\}$. The measurement matrix $\bPhi$ is a $\pm 1$ random matrix ($\bPhi\in\mathbb{R}^{m\times n}$). And each column of it is normalized to have a unit $l_2$-norm.

For better observing the performance of the proposed algorithm, several $l_1$-norm based sparse approximation algorithms are implemented. They are L1Magic package \cite{candes2005l1}, SPGL1 package \cite{spgl12007}, homotopy method \cite{malioutov2005homotopy}, the improved PNN-based sparse reconstruction (IPNNSR) method \cite{liu2016l1} and the LPNN-LCA method given by \eqref{eq-dynamics2}.
The L1Magic is a toolbox which can be used to solve the BP problem with the primal-dual interior point method. The SPGL1 package is also a common way for solving the BP problem, it is mainly based on the spectral projected gradient (SPG) method which uses the gradient vector as a search direction and chooses its step size according to the spectrum of the underlying local Hessian.
Homotopy is a kind of greedy iterative algorithm. IPNNSR and LPNN-LCA are both analog neural networks. The first one is based on the projection neural network, and its global convergence has been proved \cite{liu2016l1}. While the second one is our previous work which is based on LPNN and LCA, and it only has local asymptotic convergence.

\begin{figure}[!h]
\centering
\subfloat[]{
\includegraphics[height=1.9in]{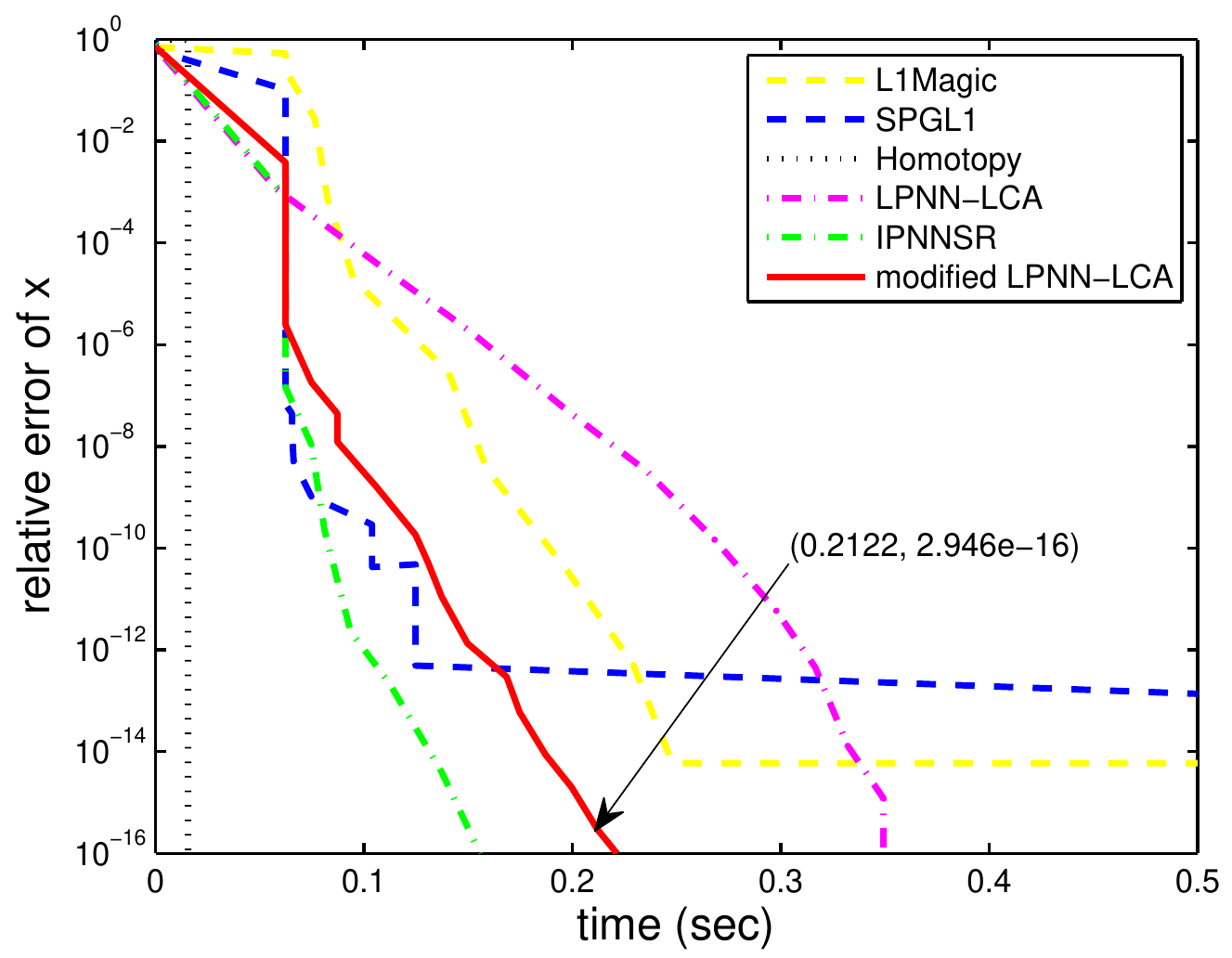}}
\hfil
\subfloat[]{
\includegraphics[height=1.9in]{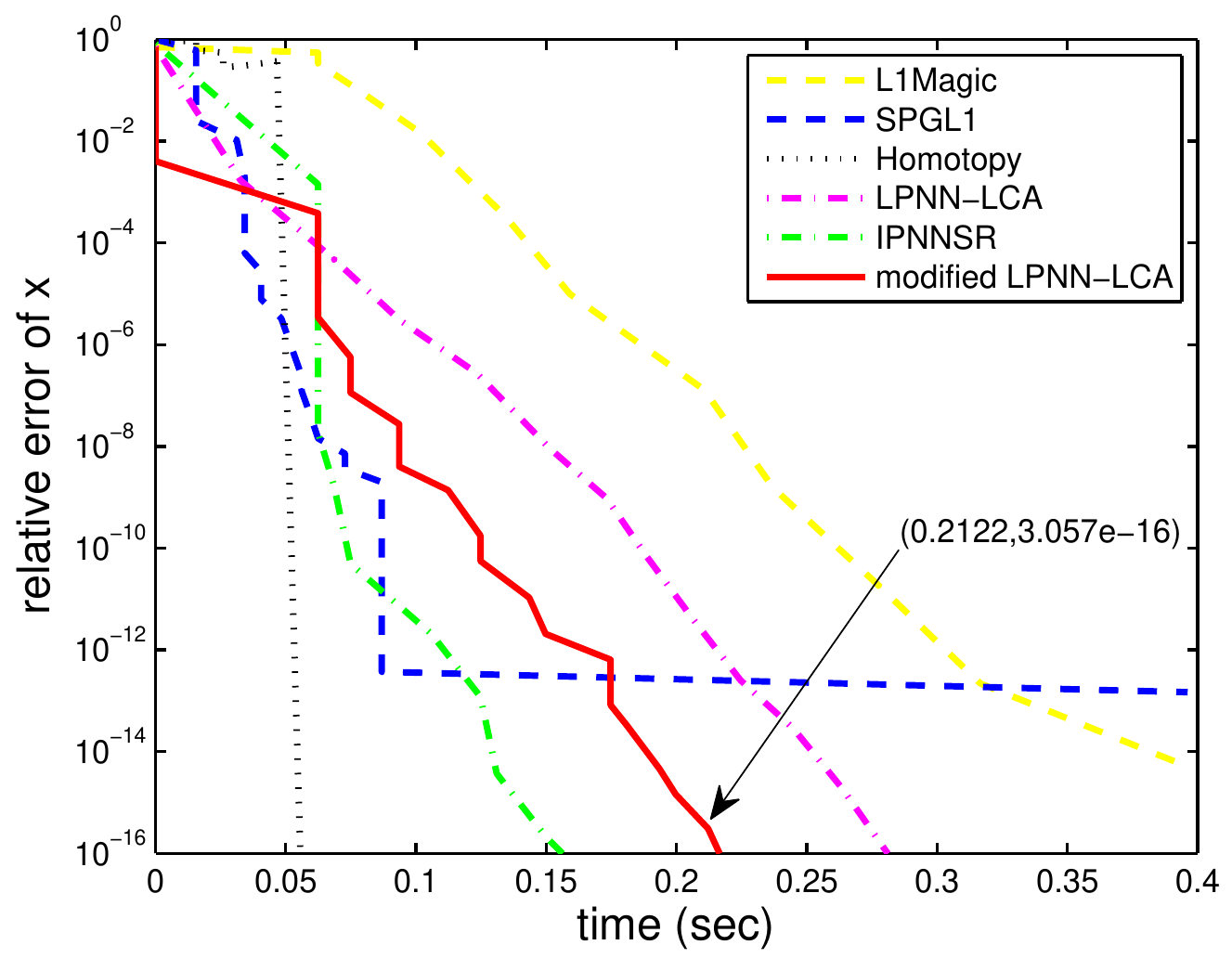}}
\hfil
\subfloat[]{
\includegraphics[height=1.9in]{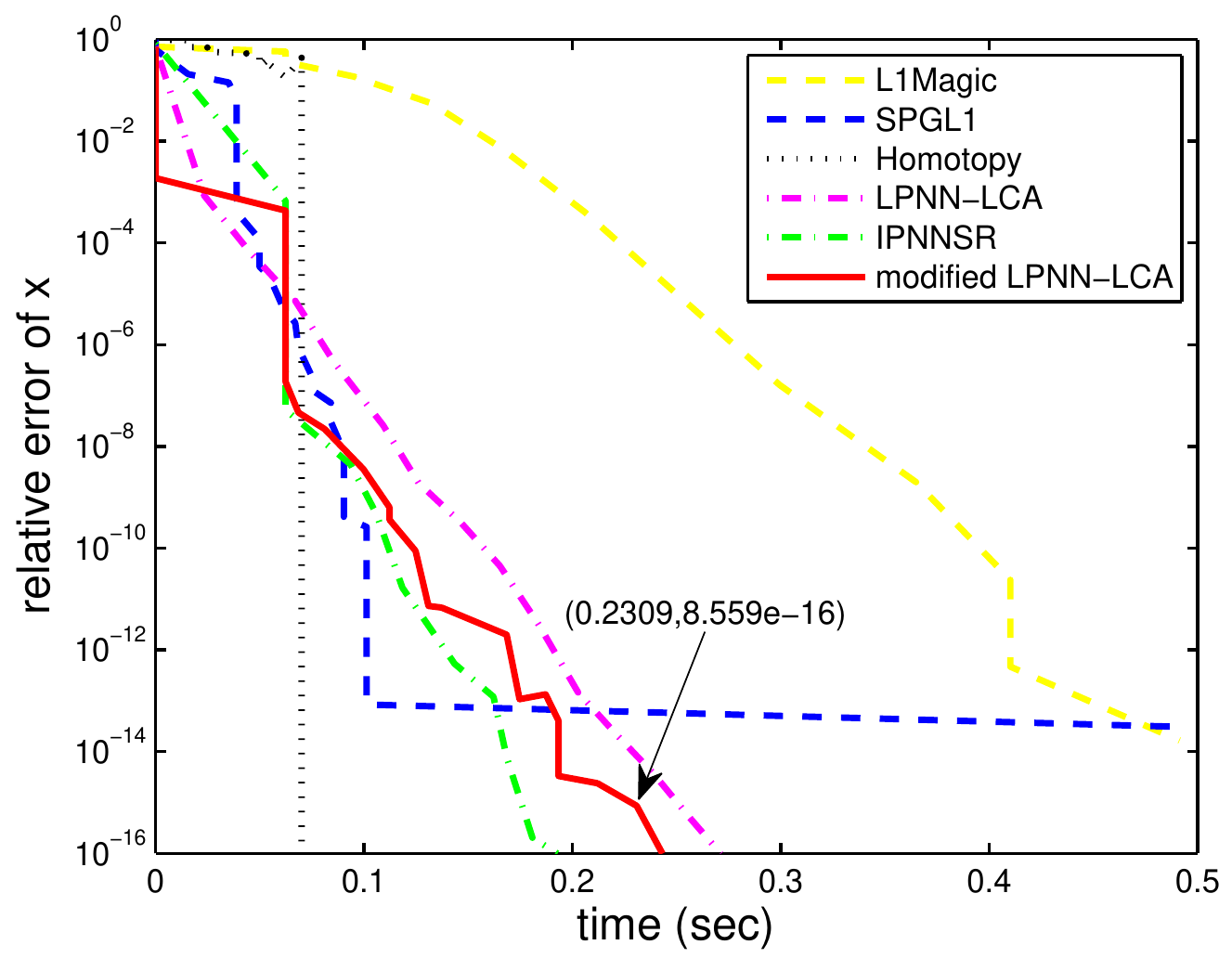}}
\caption{Relative error of $\ibx$ with respect to the convergence time. (a) $n=512$, $m=150$, and $\Omega=15$. (b) $n=512$, $m=150$, and $\Omega=20$. (c) $n=512$, $m=150$, and $\Omega=25$.}.
\label{fig:time512}
\end{figure}

\begin{figure}[!h]
\centering
\subfloat[]{
\includegraphics[height=1.9in]{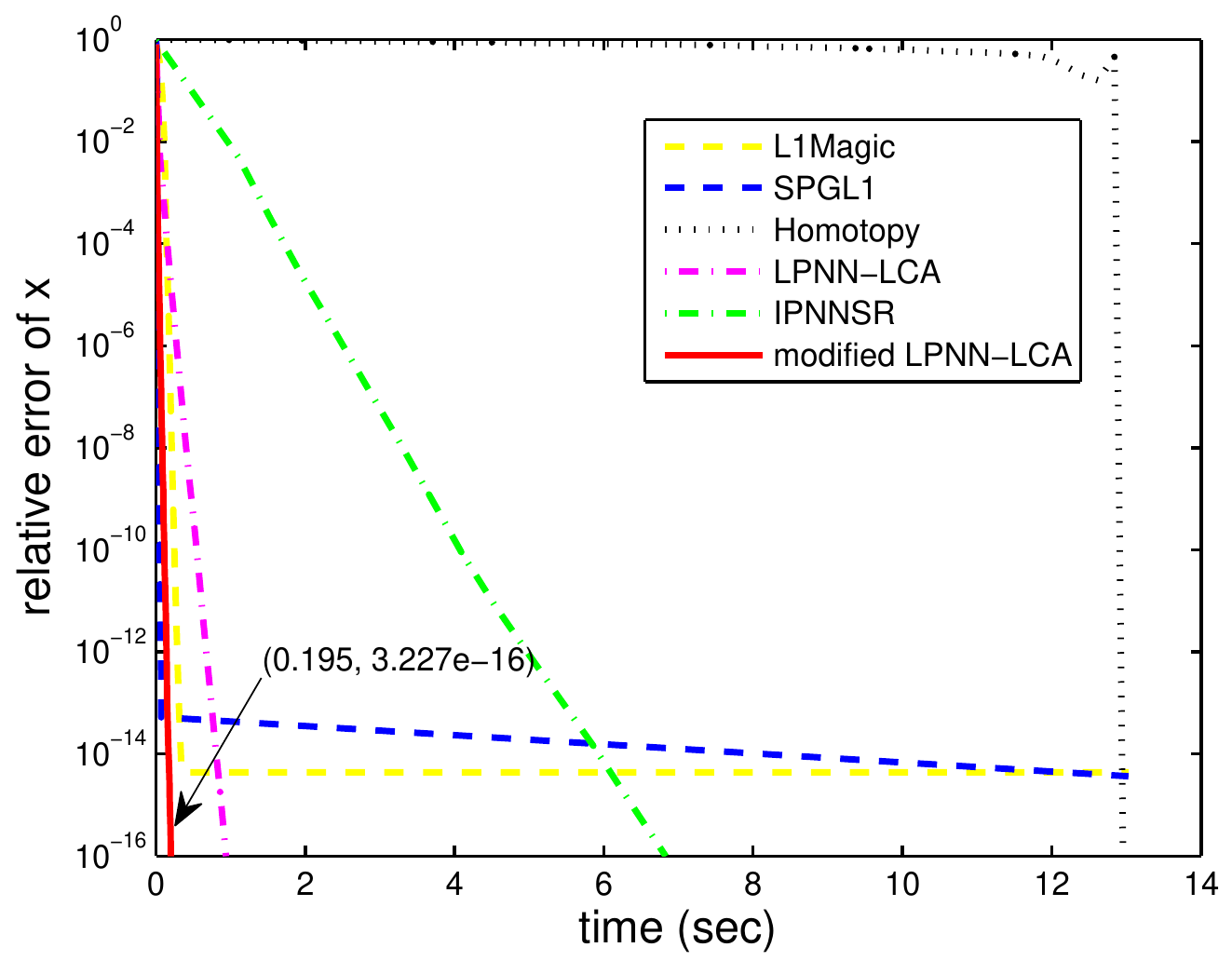}}
\hfil
\subfloat[]{
\includegraphics[height=1.9in]{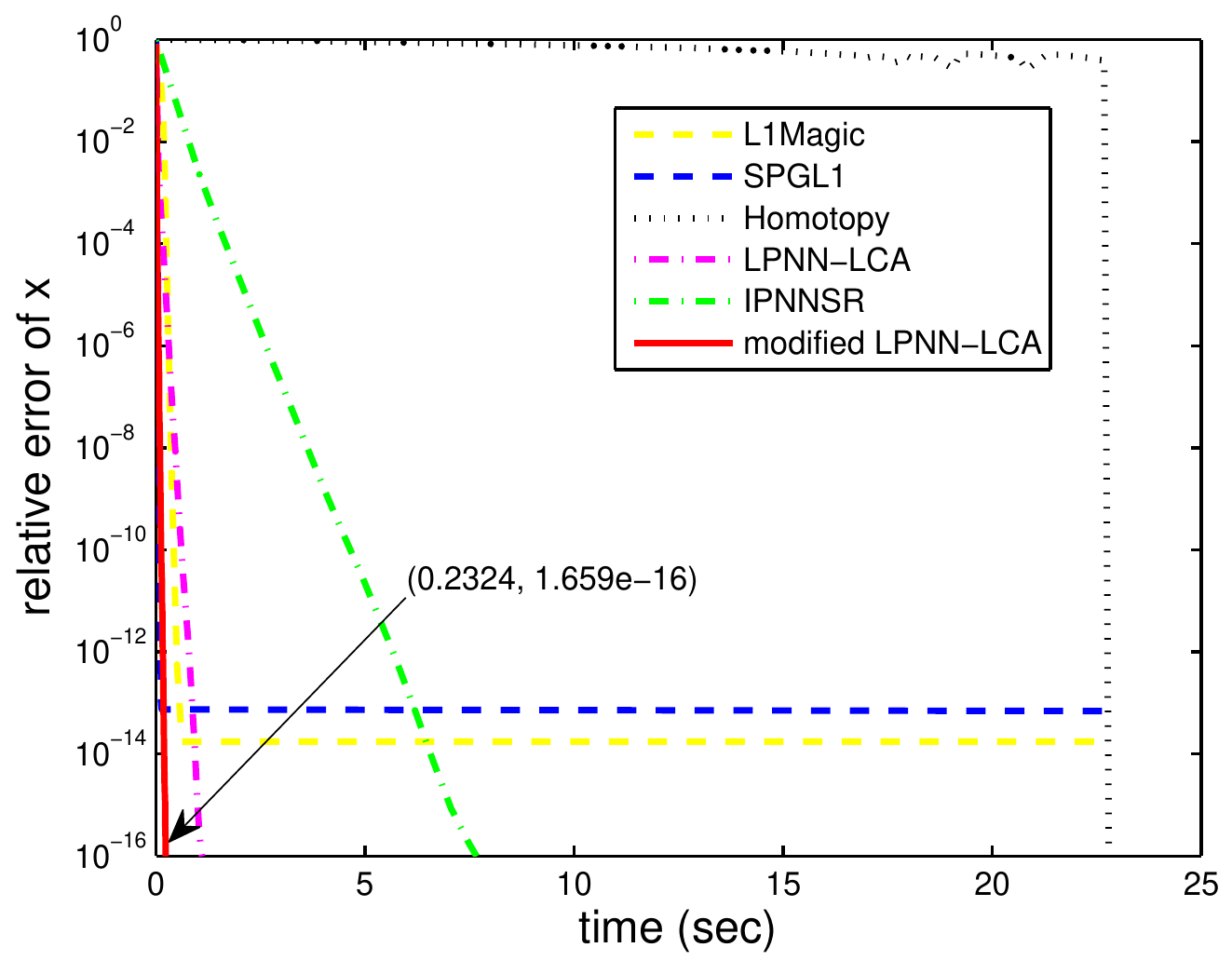}}
\hfil
\subfloat[]{
\includegraphics[height=1.9in]{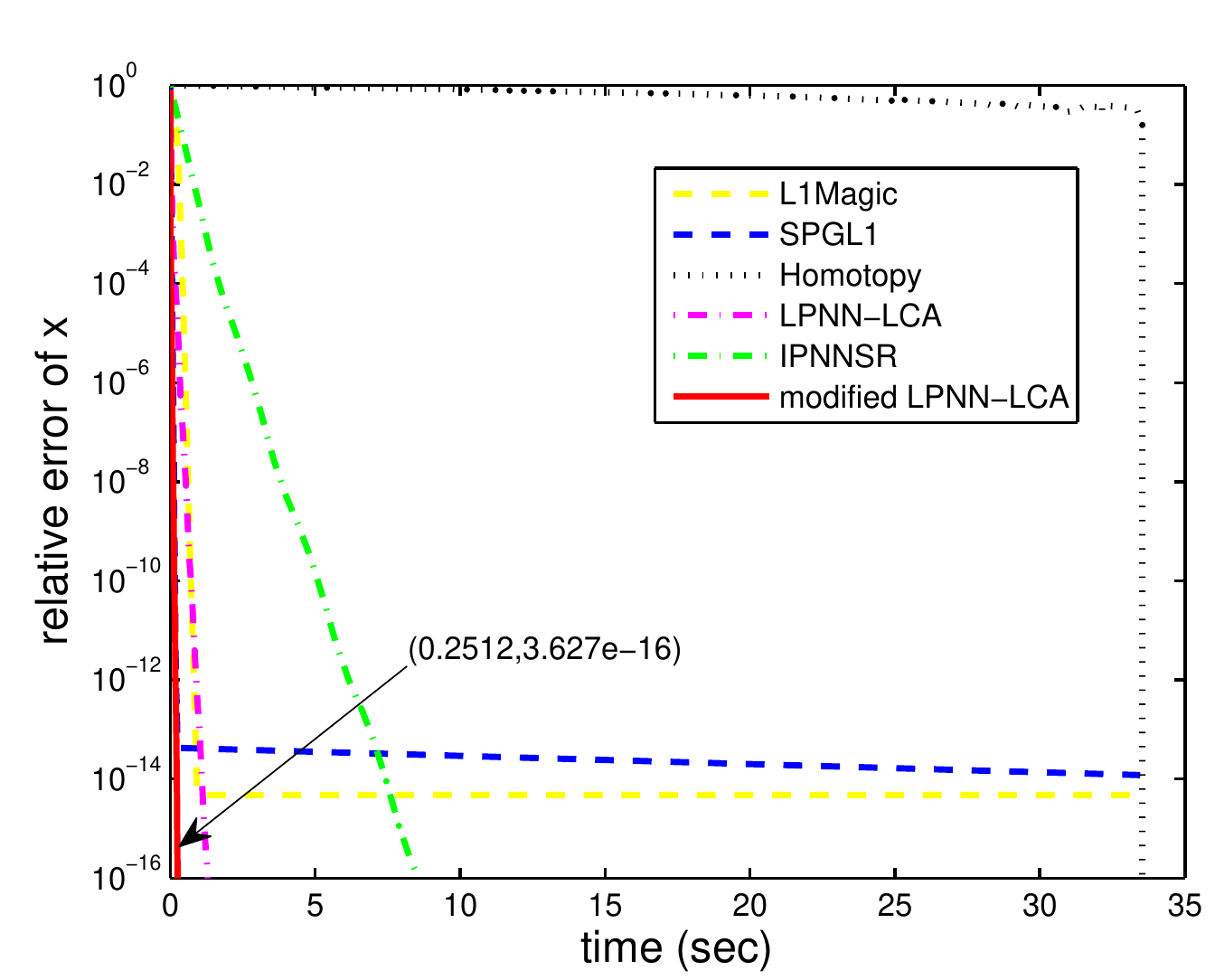}}
\caption{Relative error of $\ibx$ with respect to the convergence time. (a) $n=4096$, $m=800$, and $\Omega=75$. (b) $n=4096$, $m=800$, and $\Omega=100$. (c) $n=4096$, $m=800$, and $\Omega=125$.}.
\label{fig:time4096}
\end{figure}

\begin{figure*}[!h]
\centering
\begin{tabular}{c@{\extracolsep{2mm}}c@{\extracolsep{2mm}}c}
\mbox{\epsfig{figure=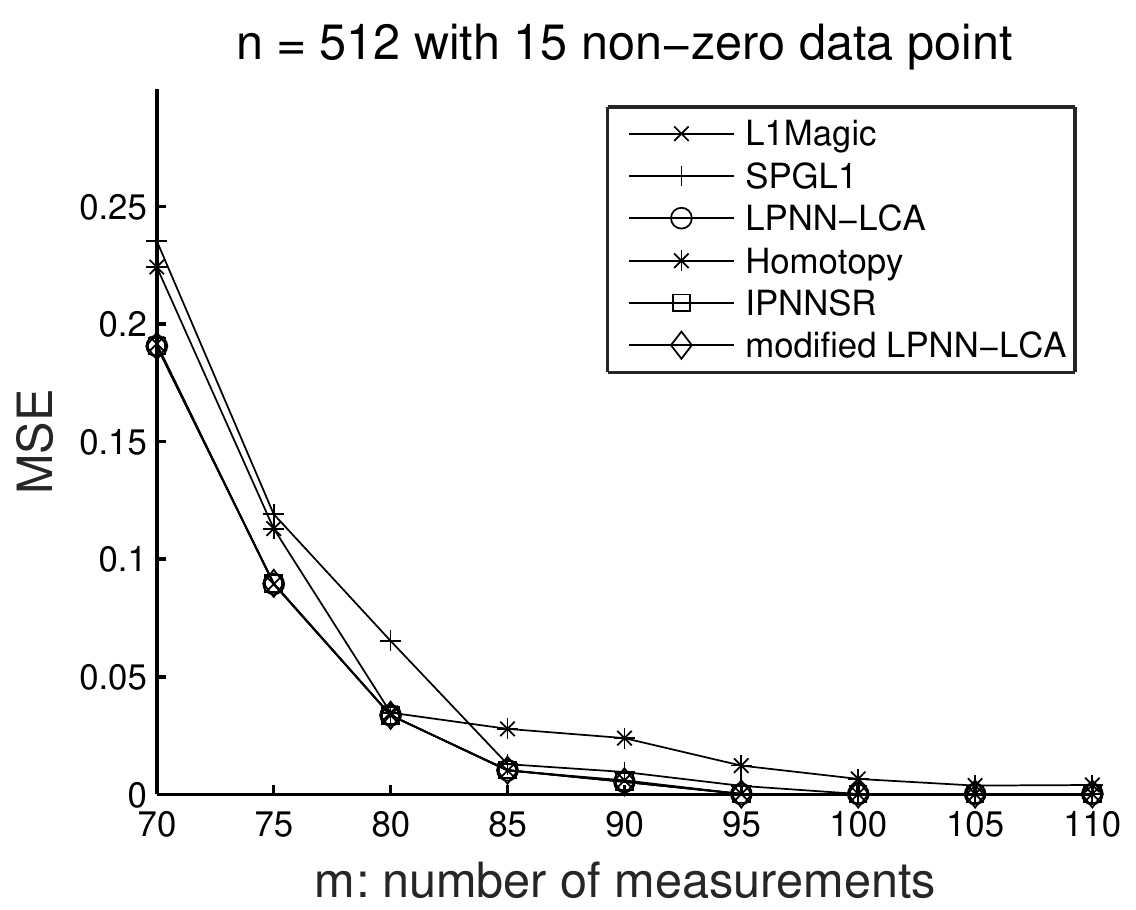,width=2.1in}} &
\mbox{\epsfig{figure=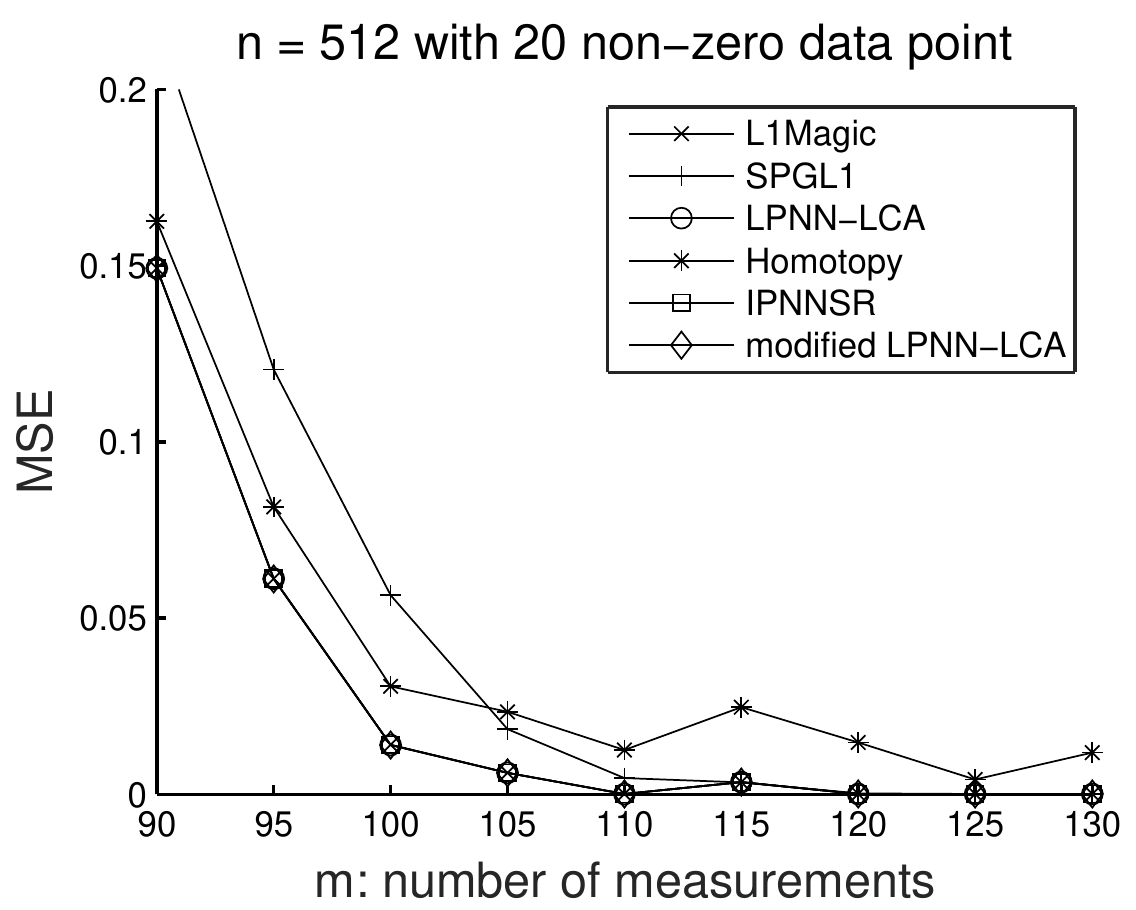,width=2.1in}} &
\mbox{\epsfig{figure=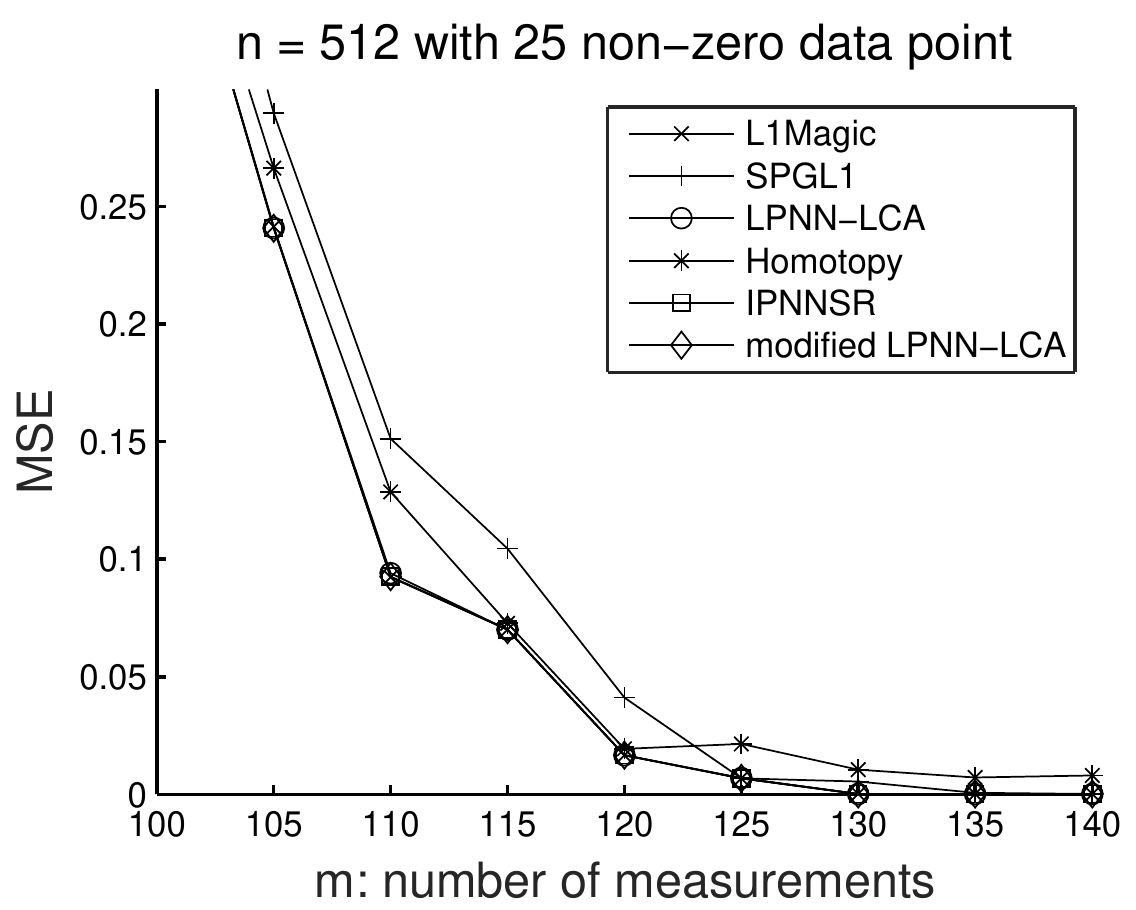,width=2.1in}}\\
(a)&(b)&(c)\\
\mbox{\epsfig{figure=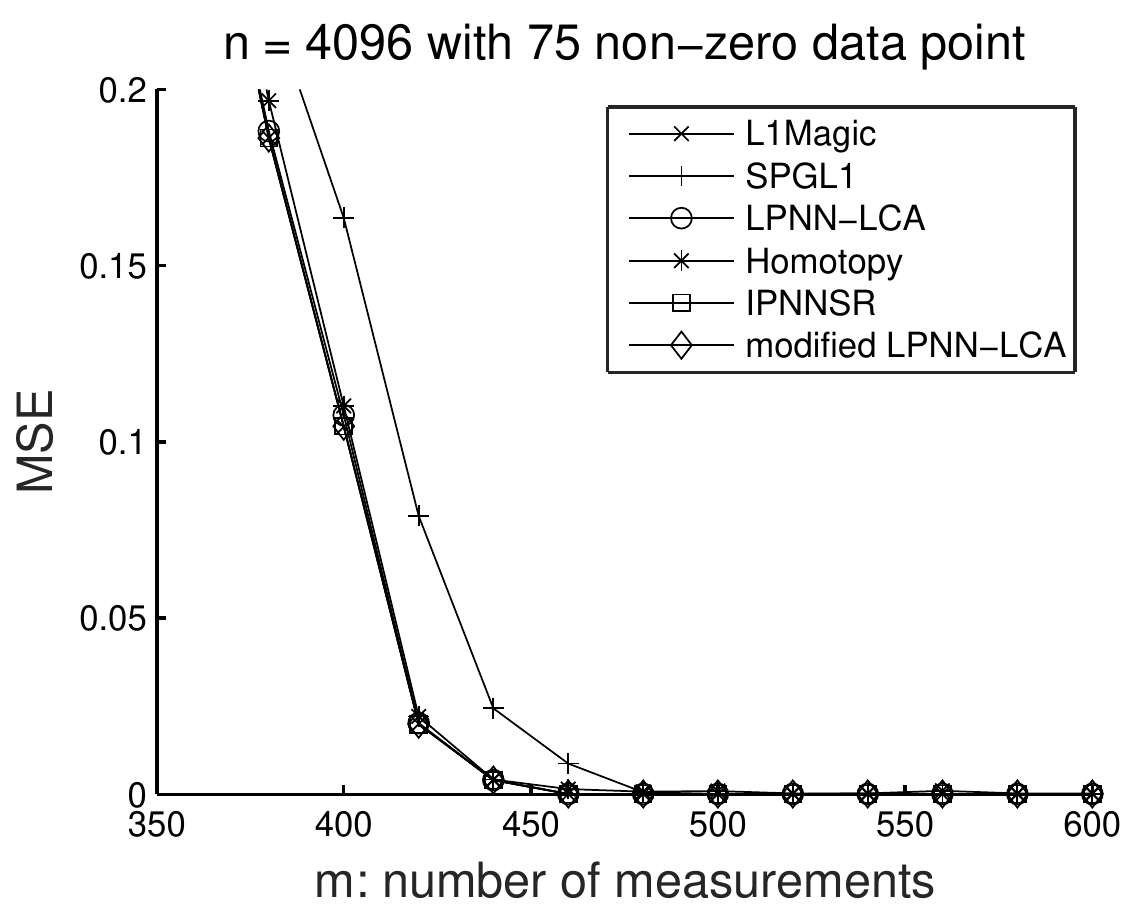,width=2.1in}} &
\mbox{\epsfig{figure=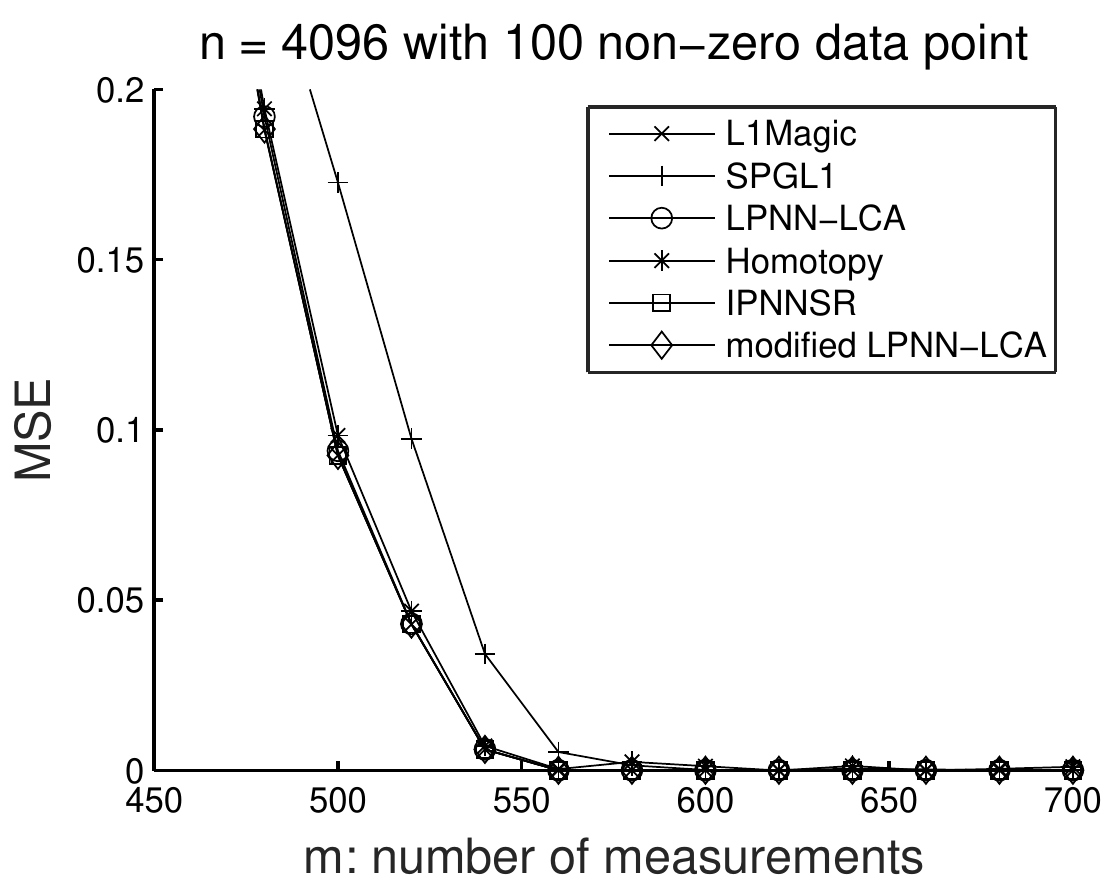,width=2.1in}} &
\mbox{\epsfig{figure=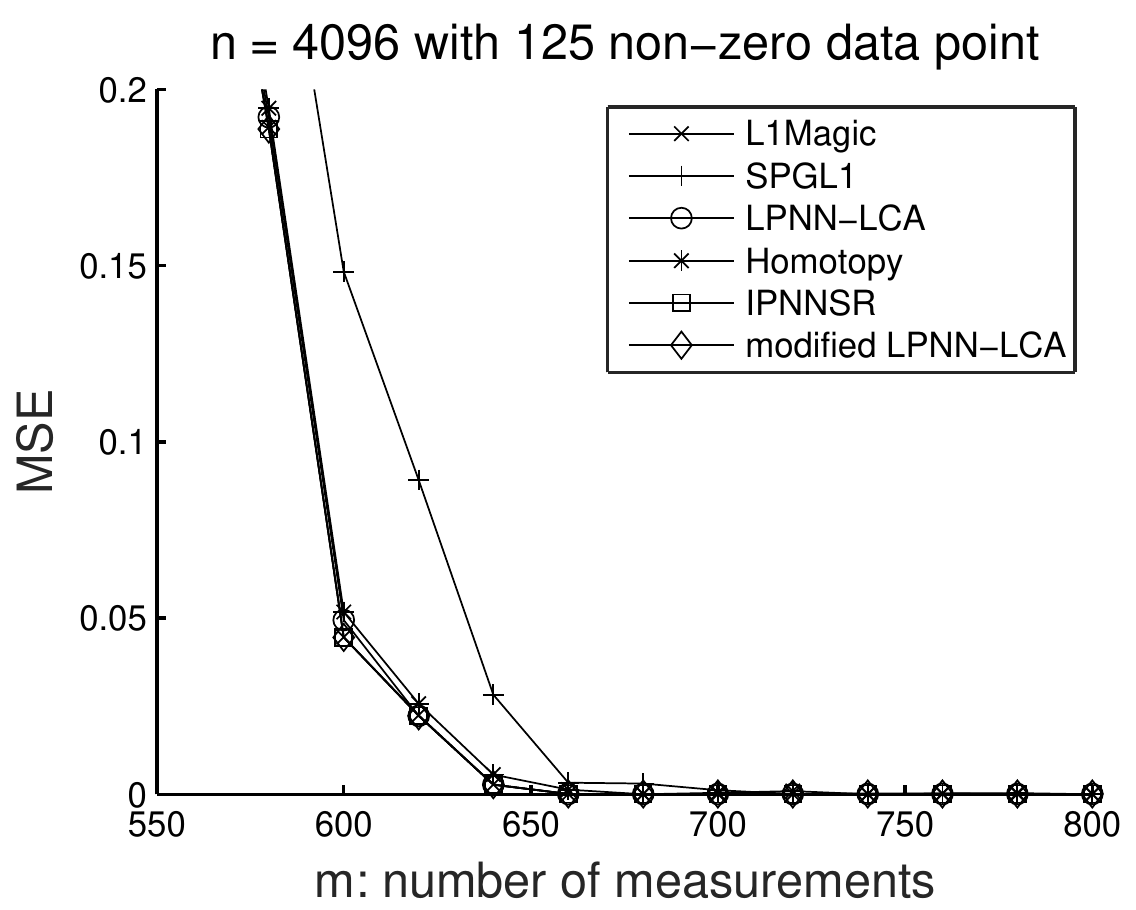,width=2.1in}}\\
(d)&(e)&(f)\\
\end{tabular}
\caption{MSE performances among the six methods. (a) $n=512$ and $\Omega=15$. (b) $n=512$ and $\Omega=20$. (c) $n=512$ and $\Omega=25$. (d) $n=4096$ and $\Omega=75$. (e) $n=4096$ and $\Omega=100$. (f) $n=4096$ and $\Omega=125$.}
\label{fig:MSE1}
\end{figure*}

\begin{figure*}[!h]
\centering
\begin{tabular}{c@{\extracolsep{2mm}}c@{\extracolsep{2mm}}c}
\mbox{\epsfig{figure=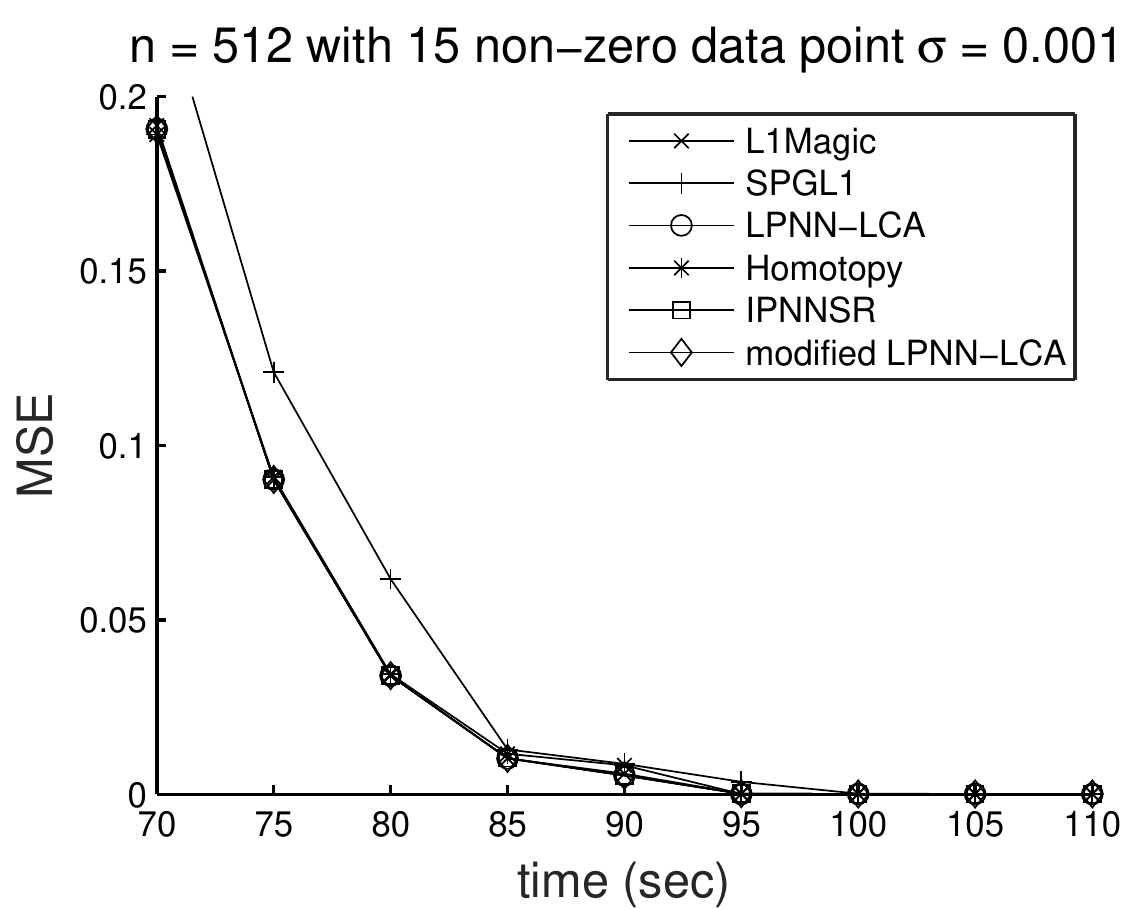,width=2.1in}} &
\mbox{\epsfig{figure=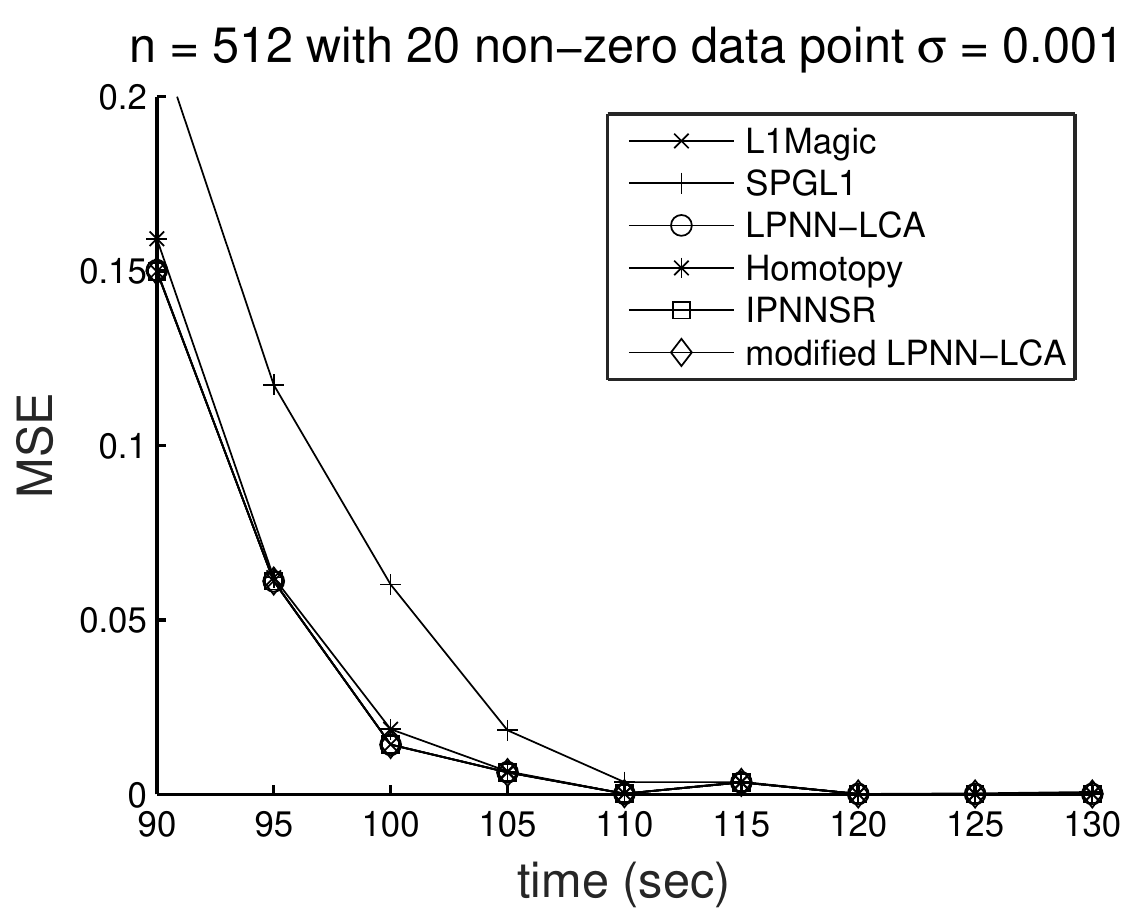,width=2.1in}} &
\mbox{\epsfig{figure=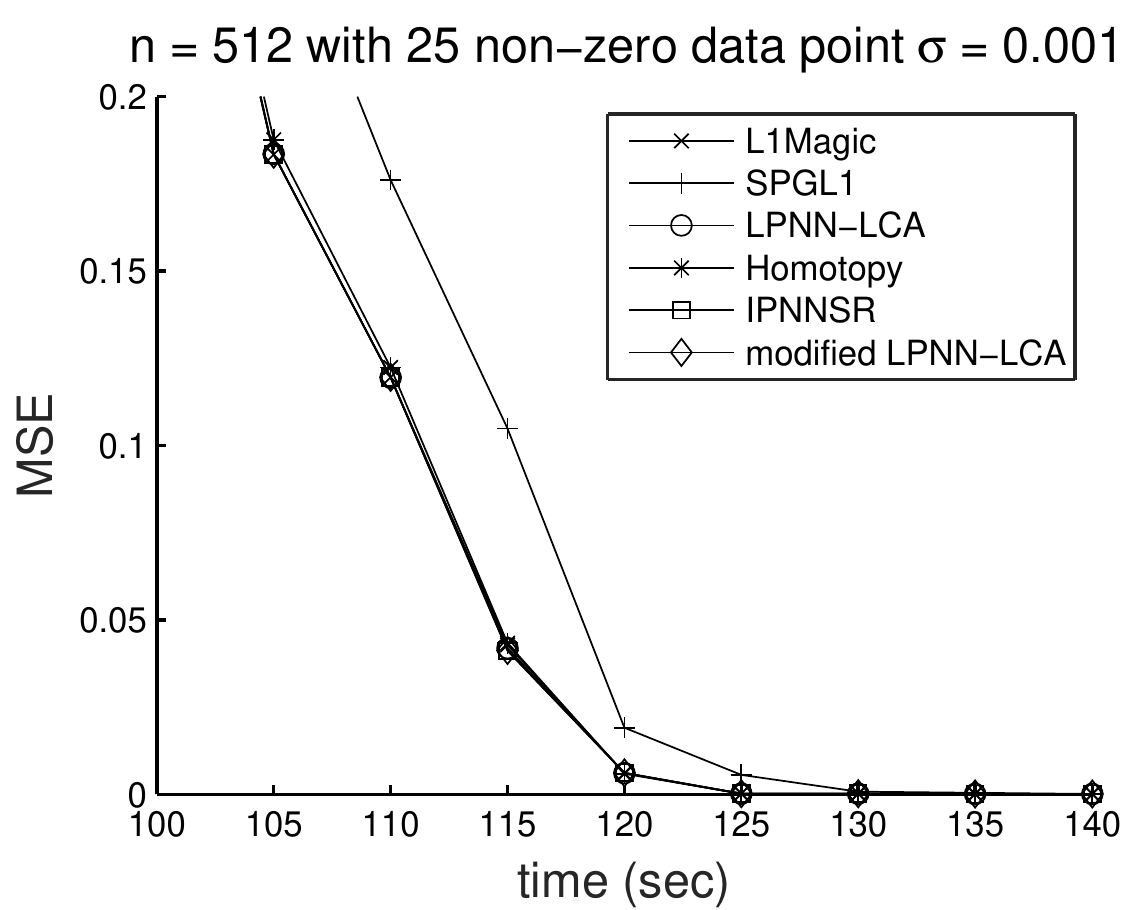,width=2.1in}}\\
(a)&(b)&(c)\\
\mbox{\epsfig{figure=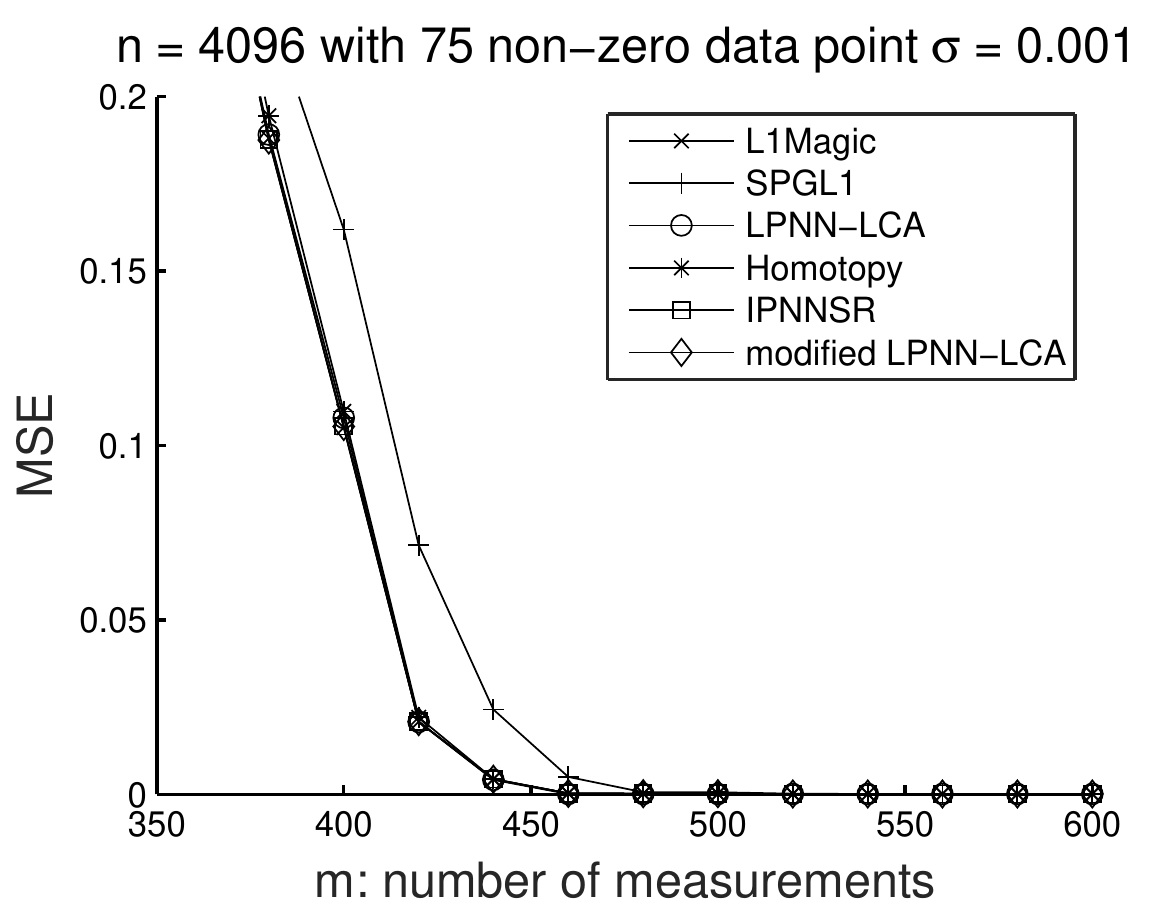,width=2.1in}} &
\mbox{\epsfig{figure=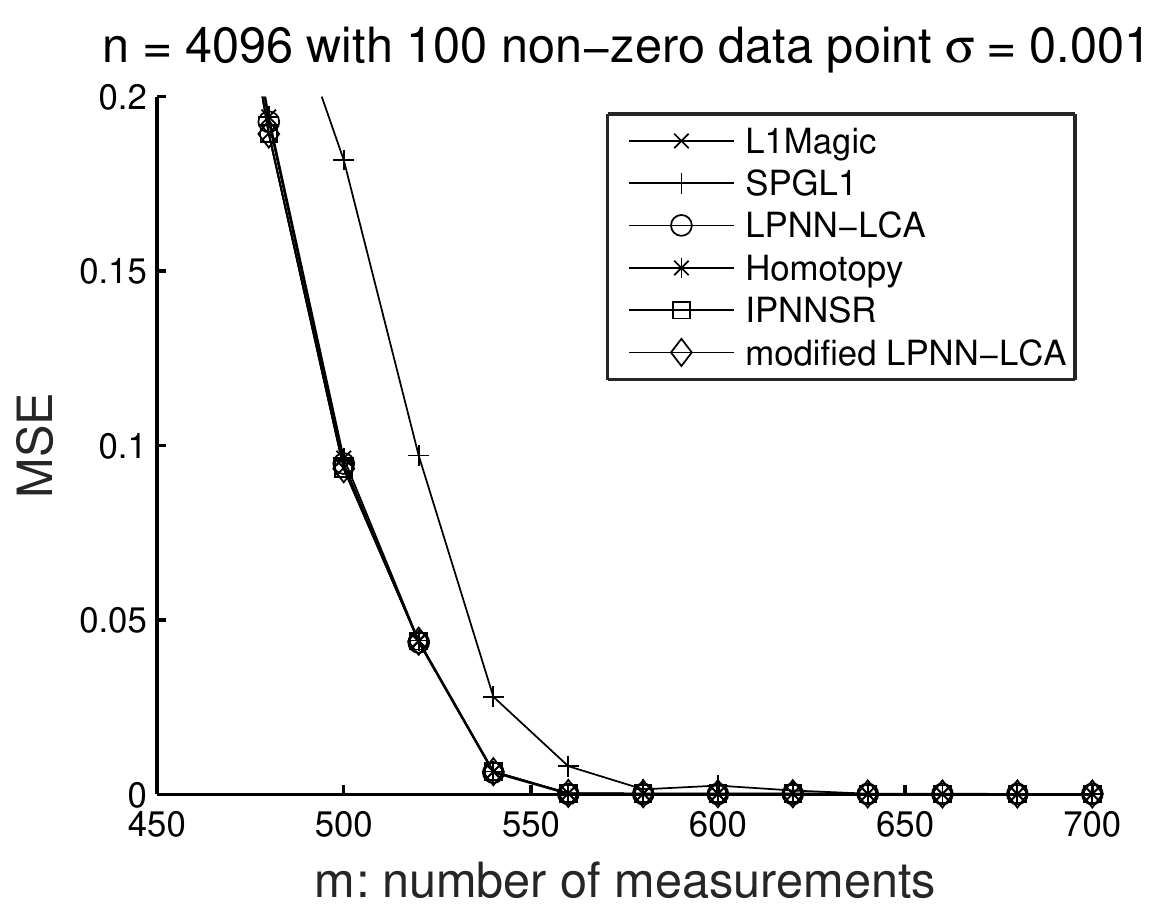,width=2.1in}} &
\mbox{\epsfig{figure=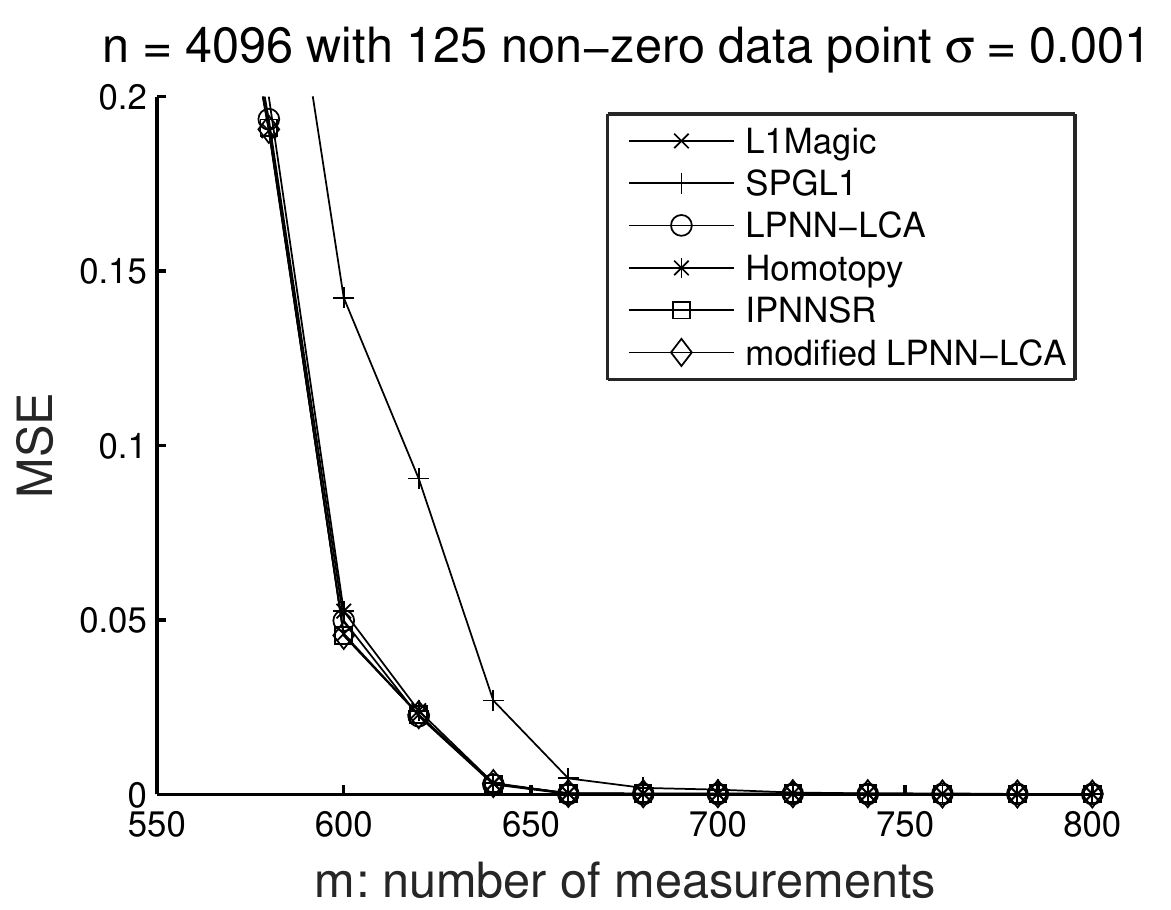,width=2.1in}}\\
(d)&(e)&(f)\\
\end{tabular}
\caption{MSE performances under Gaussian noise, where $\sigma=0.001$. (a) $n=512$ and $\Omega=15$. (b) $n=512$ and $\Omega=20$. (c) $n=512$ and $\Omega=25$. (d) $n=4096$ and $\Omega=75$. (e) $n=4096$ and $\Omega=100$. (f) $n=4096$ and $\Omega=125$.}
\label{fig:MSE2}
\end{figure*}

\begin{figure*}[!h]
\centering
\begin{tabular}{c@{\extracolsep{2mm}}c@{\extracolsep{2mm}}c}
\mbox{\epsfig{figure=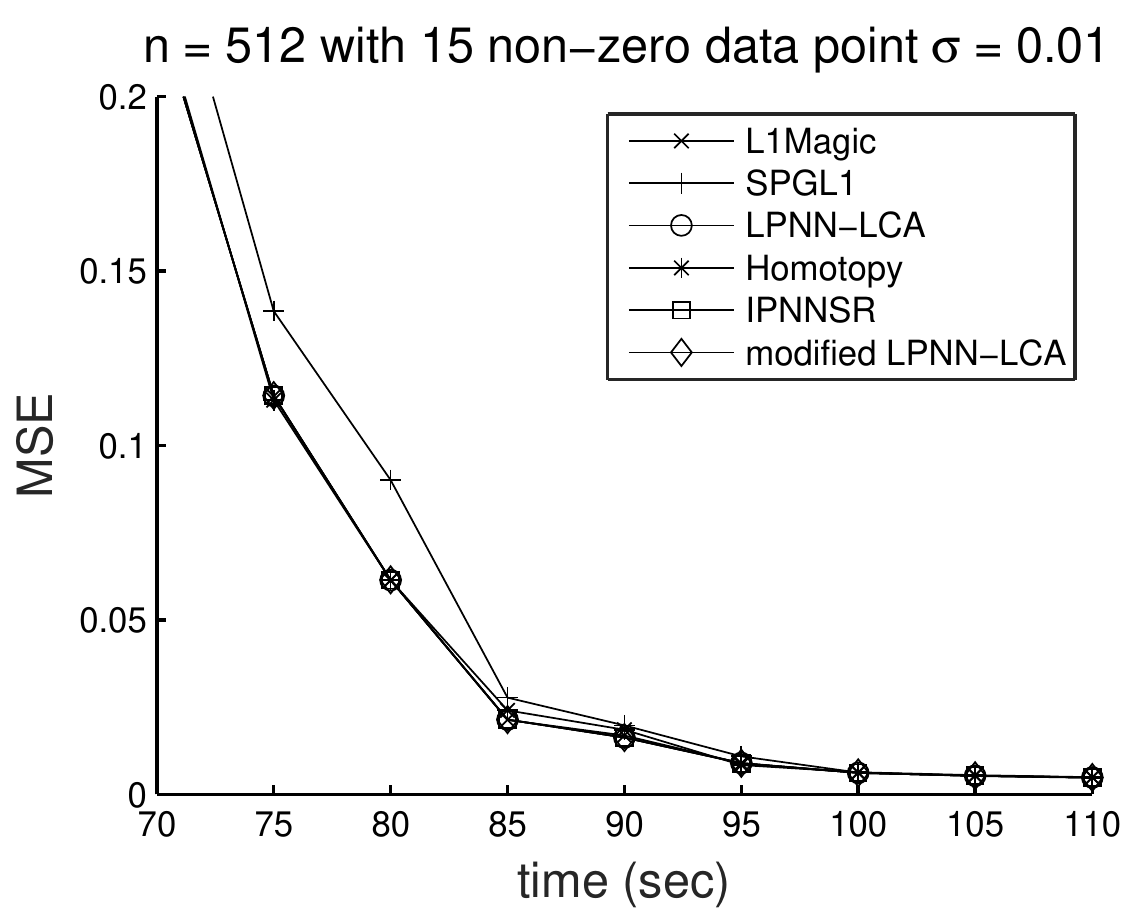,width=2.1in}} &
\mbox{\epsfig{figure=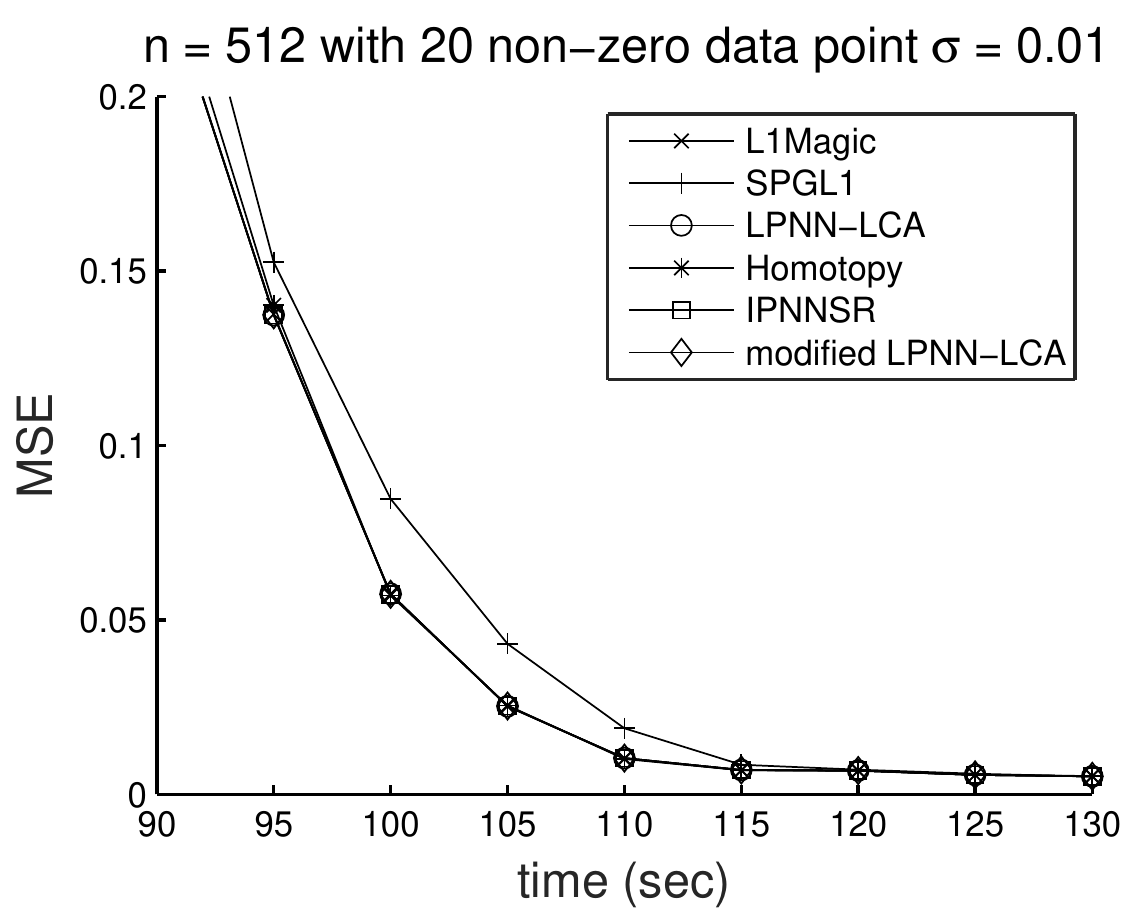,width=2.1in}} &
\mbox{\epsfig{figure=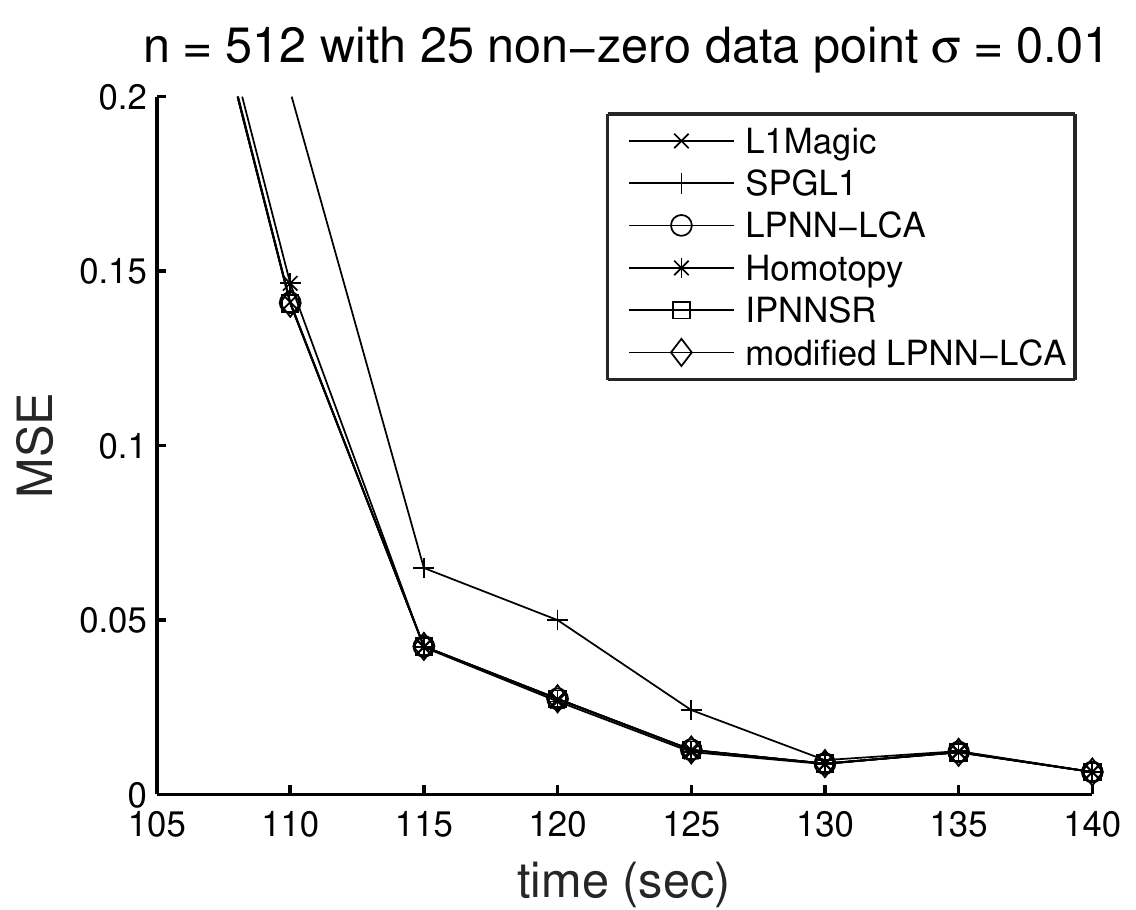,width=2.1in}}\\
(a)&(b)&(c)\\
\mbox{\epsfig{figure=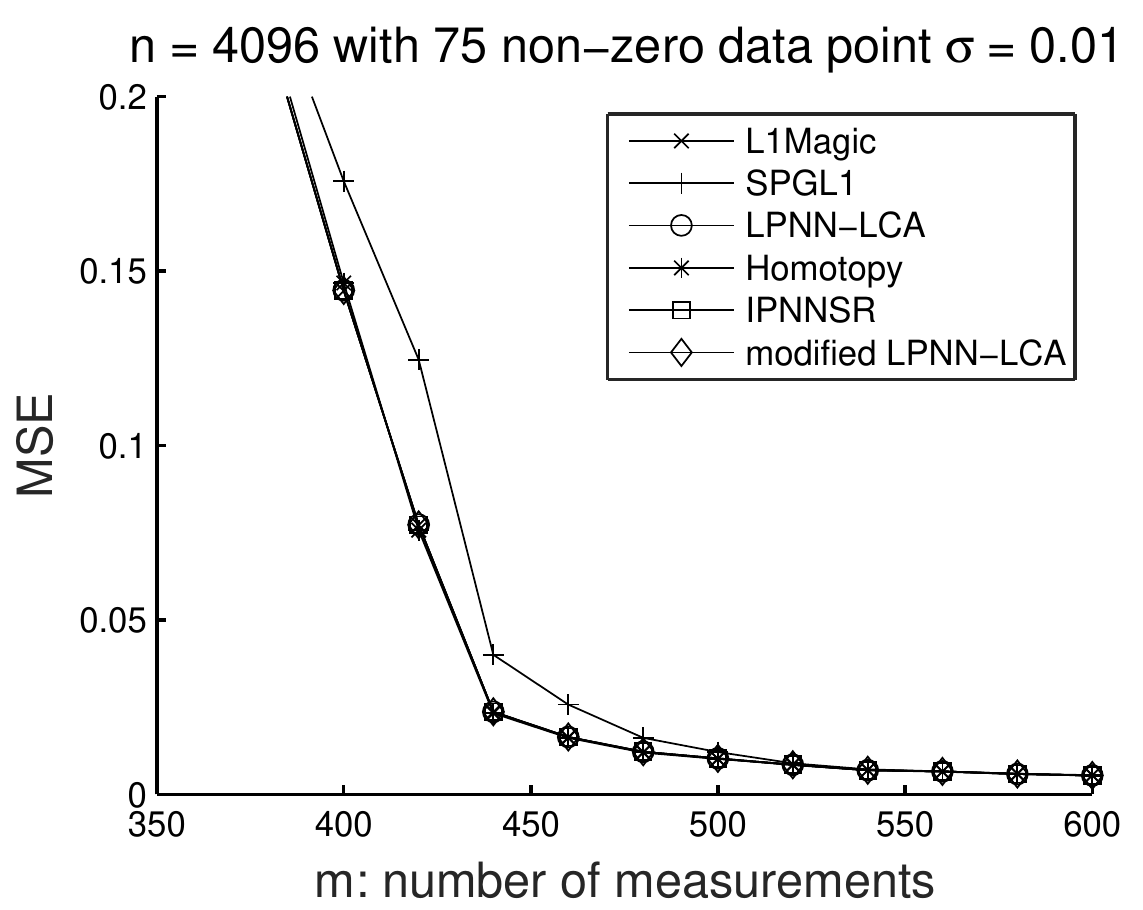,width=2.1in}} &
\mbox{\epsfig{figure=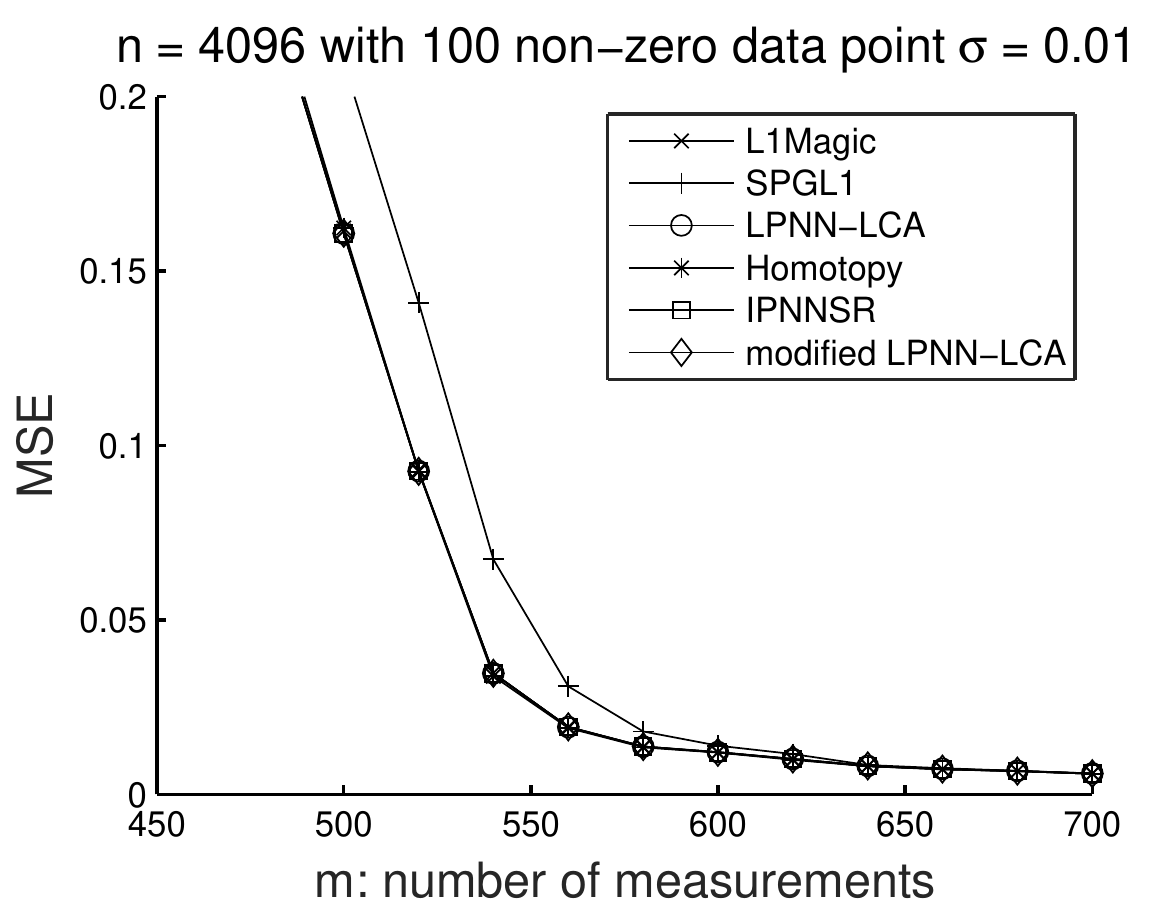,width=2.1in}} &
\mbox{\epsfig{figure=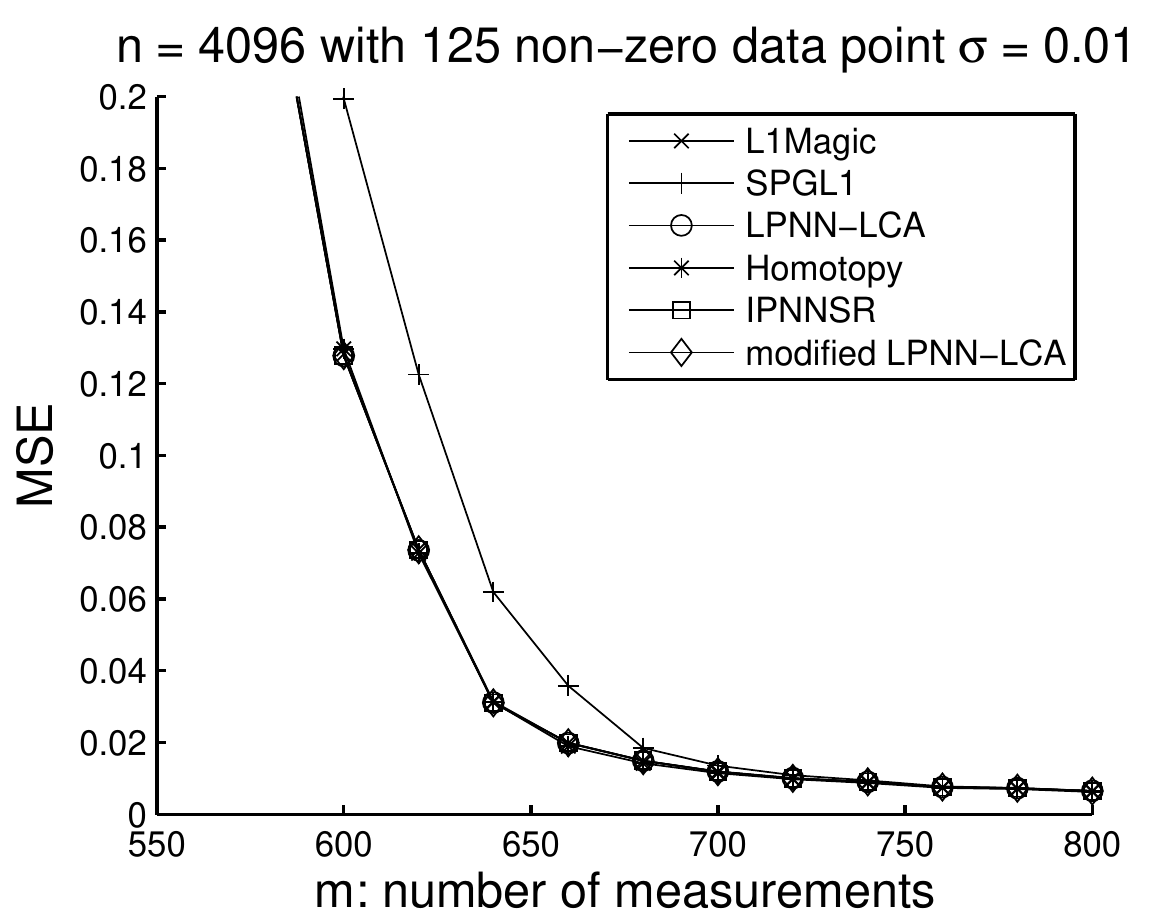,width=2.1in}}\\
(d)&(e)&(f)\\
\end{tabular}
\caption{MSE performances under Gaussian noise, where $\sigma=0.01$. (a) $n=512$ and $\Omega=15$. (b) $n=512$ and $\Omega=20$. (c) $n=512$ and $\Omega=25$. (d) $n=4096$ and $\Omega=75$. (e) $n=4096$ and $\Omega=100$. (f) $n=4096$ and $\Omega=125$.}
\label{fig:MSE3}
\end{figure*}

\subsection{Stability}
In Section~\ref{section4}, we theoretically prove the global convergence of the proposed method. Then, we use several typical examples to show its convergence in practice. For both Fig.~\ref{fig:convergence512} and Fig.~\ref{fig:convergence4096}, the first three columns are the dynamics of the estimated parameters $\ibx$, $\ibu$, and $\blam$ and the corresponding recovered signals are shown in the last column. From Fig.~\ref{fig:convergence512}, it is observed that when $n=512$, the dynamics can settle down within around 60-80 iterations. While for $n=4096$ in Fig.~\ref{fig:convergence4096}, the dynamics can settle down within 100 to 120 iterations. From these figures, we know that the whole system is stable and can converge to their optimal solutions.

\subsection{Convergence time}
In this experiment, we test the speed of the proposed method. First let $n=512$, $m=150$, and $\Omega={15,20,25}$. The results are shown in Fig.~\ref{fig:time512}. Where the y-axis denotes the average relative error of ten trials, and the x-axis is the CPU time (unit: sec). While Fig.~\ref{fig:time4096} depicts the results when $n=4096$, $m=800$, and $\Omega={75,100,125}$. From Fig.~\ref{fig:time512} and Fig.~\ref{fig:time4096}, it is observed that, when $n=512$, the greedy method homotopy is better than others. And compared with other two analog neural networks, the IPNNSR needs less convergence time to achieve a satisfied result.
When $n=4096$, the performance of our proposed algorithm is superior. Comparing with others, the proposed algorithm has relatively stable convergence time.

Besides, it is worth noting that these three analog neural networks can be implemented with hardware circuit, hence their speed can be further improved. 
The computational complexity of IPNNSR's dynamics is ${\cal O}(n^2)$, while, for the proposed algorithm and LPNN-LCA, their corresponding value is ${\cal O}(mn)$. Hence, compared with the improved LPNN-LCA and the original LPNN-LCA, the increase of $n$ has a greater impact on IPNNSR. Comparing with LPNN-LCA, the improved method has better stability. And its step size $\mu=1$ in the experiments. However, the step size of the original LPNN-LCA can only be 0.1 under the same situation, otherwise the system may not converge. So the convergence time of the proposed method is less than the original LPNN-LCA framework.

\subsection{MSE performance}
In this experiment, we fix the length of the signal and the number of its nonzero elements. The MSE performances of the proposed method are investigated with different $m$. We repeat the experiment $100$ times with different measurement matrices, initial states and sparse signals. The results are given in Fig.~\ref{fig:MSE1}.
Where the x-axis denotes the number of elements in observation vector $\ibb$, the y-axis is the mean square error (MSE) between the recovered signal and the original one. It can be seen from the figures that as $m$ increases, the MSEs of all algorithms decrease.
In the same situation, the SPGL1 package and the homotopy method have larger MSEs. For other four approaches, their MSE performance is quite similar with each other.

Even the proposed method is devised for solving the BP problem, it still works
under low level Gaussian noise. The corresponding simulation results are given in Fig.\ref{fig:MSE2} and Fig.\ref{fig:MSE3}. Their observations are generated by
\beq
\ibb=\bPhi\ibx+\beps, \nonumber
\eeq
where $\beps=[\eps_1, ..., \eps_m]^\mathrm{T}$ denotes the vector of zero-mean Gaussian noise. The standard deviation of the noise is denotes by $\sigma$. For the Fig.\ref{fig:MSE2}, $\sigma=0.001$, while for Fig.\ref{fig:MSE3}, $\sigma=0.01$.

\section{Conclusion}\label{section6}
This paper proposes a novel sparse approximation algorithm by combining LPNN, LCA and projection theorem. The main target of the proposed algorithm is to calculate a real-time solution of the large-scale problem. For this purpose, this paper uses LPNN framework and introduces LCA to set up an approximate differentiable expression for the sub-differential at the non-differentiable point. Referring to the dynamic of projection neural network, we further modify its dynamics. Thus the equilibrium point of the improved dynamics is equivalent to an optimal solution, and the proposed algorithm is globally stable in the sense of Lyapunov. Besides, according to the simulation results, the MSE performance of the proposed algorithm is similar with several state-of-the-art methods and for large-scale problem it has obvious advantage in speed.

In future work we will try to modify the dynamics of LPNN-LCA method for solving BPDN problem and prove its global convergence. Besides, the modified dynamics may be helpful to improve the performance of the approximate $l_0$-norm relevant methods.

\appendices

\ifCLASSOPTIONcaptionsoff
  \newpage
\fi



%
%
\bibliographystyle{IEEEtran}
\bibliography{MLPNN_reference}

%

\begin{IEEEbiography}[{\includegraphics[width=1in,height=1.25in,clip,keepaspectratio]{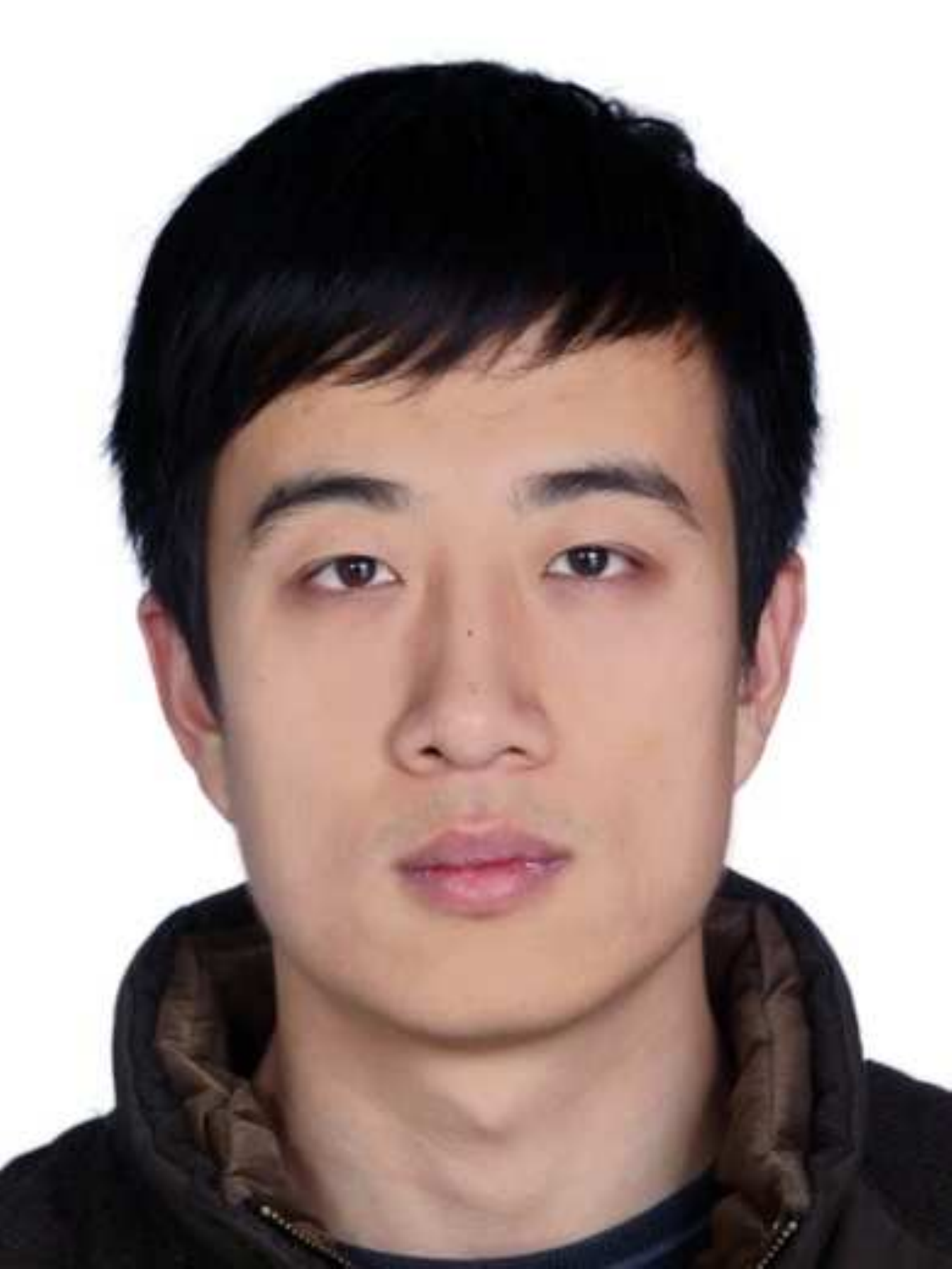}}]{Hao Wang}
is currently pursuing Ph.D. degree from the Department of Electronic Engineering, City University of Hong Kong. His current research interests include neural networks and machine learning.
\end{IEEEbiography}

\begin{IEEEbiography}[{\includegraphics[width=1in,height=1.25in,clip,keepaspectratio]{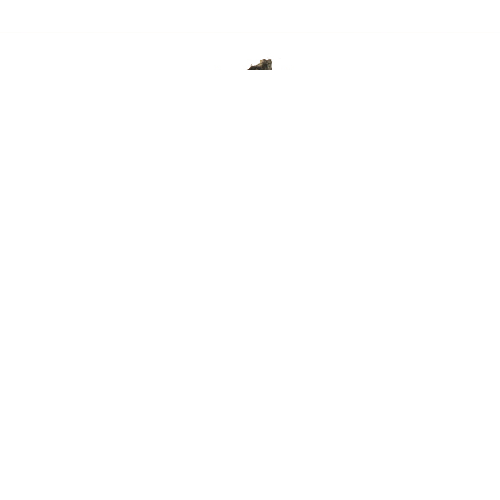}}]{Ruibin Feng}
received the PhD. degree in electronic engineering from the City University of Hong Kong in 2017.
His current research interests include Neural Networks and Machine Learning.
\end{IEEEbiography}

\begin{IEEEbiography}[{\includegraphics[width=1in,height=1.25in,clip,keepaspectratio]{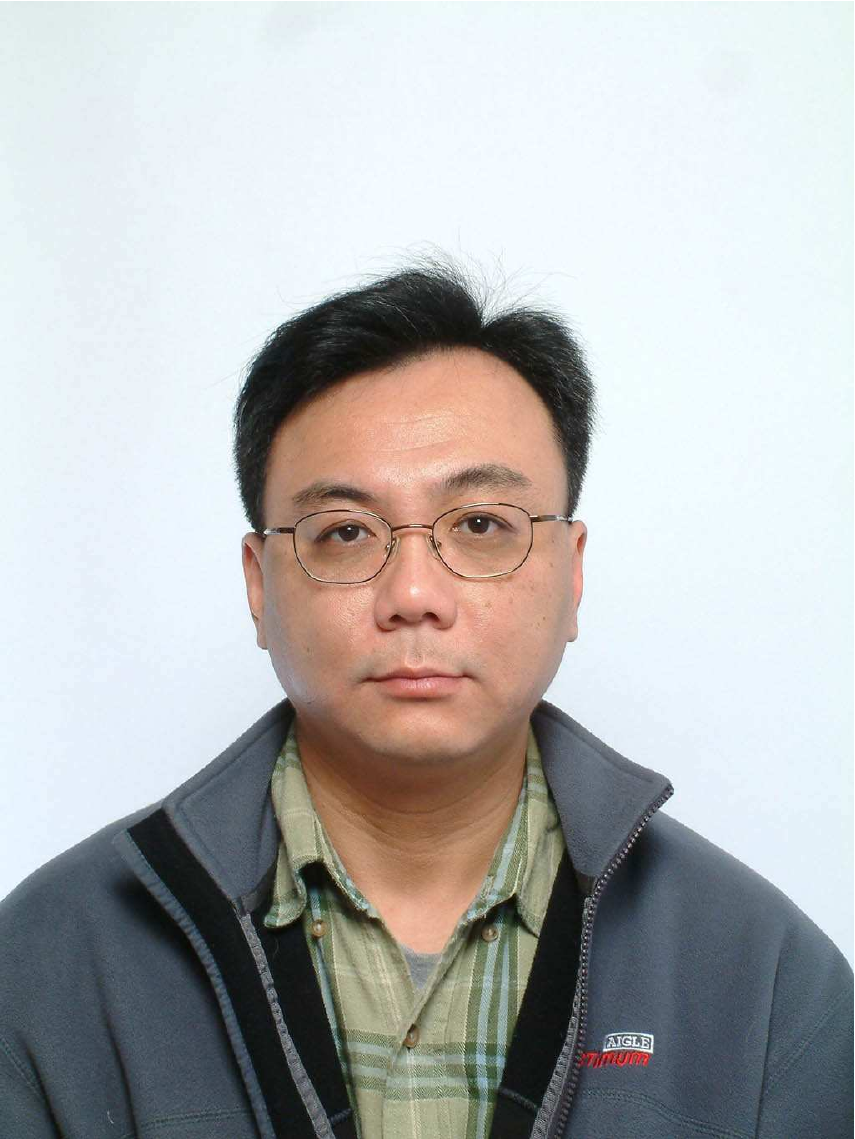}}]{Chi-Sing Leung}
received the Ph.D. degree from the Chinese University of Hong Kong, Hong Kong, in 1995.

He is currently a Professor with the Department of Electronic Engineering, City University of
Hong Kong, Hong Kong. He has authored over 120 journal papers in the areas of digital signal
processing, neural networks, and computer graphics. His current research interests include neural computing and computer graphics.

Dr. Leung was a member of the Organizing Committee of ICONIP2006. He received the 2005 IEEE Transactions on Multimedia Prize Paper Award for his paper titled The Plenoptic Illumination Function in 2005. He was the Program Chair of ICONIP2009 and ICONIP2012. He is/was the Guest Editor of several journals, including Neural Computing and Applications, Neurocomputing, and Neural Processing Letters. He is a Governing Board Member of the Asian Pacific Neural Network
Assembly (APNNA) and the Vice President of APNNA.
\end{IEEEbiography}

\end{document}